\documentclass[prd, twocolumn, superscriptaddress, nofootinbib,10pt]{revtex4}
\usepackage{graphicx}
\usepackage{dcolumn}
\usepackage{amssymb,amsmath,amsthm,mathrsfs}
\usepackage{epstopdf}
\usepackage{lipsum}
\usepackage{hyperref}
\usepackage{color}
\usepackage{soul}
\allowdisplaybreaks

 \newcommand{\be}{\begin{equation}}
        \newcommand{\ee}{\end{equation}}
    \newcommand{\bea}{\begin{eqnarray}}
        \newcommand{\eea}{\end{eqnarray}}
                         
                \newcommand{\IM}{\mathfrak{Im}}     
                   
                    \newcommand{\dt}{dt}     
        \newcommand{\dx}{dx}      

\begin{document}

\title{Higgs-curvature coupling and post-inflationary vacuum instability}

\newcommand{\addressICL}{Department of Physics, Imperial College London, SW7 2AZ, United Kingdom.}

\newcommand{\addressIFT}{Instituto de F\'isica Te\'orica UAM-CSIC, Cantoblanco, 28049 Madrid, Spain.}

\newcommand{\addressCERN}{Theoretical Physics Department, CERN, Geneva, Switzerland}

\author{Daniel G. Figueroa}
\affiliation{\addressCERN}

\author{Arttu Rajantie}
\affiliation{\addressICL}

\author{Francisco Torrent\'i\,}
\affiliation{\addressIFT}

\date{\today}

\begin{abstract}
We study the post-inflationary dynamics of the Standard Model (SM) Higgs field in the presence of a non-minimal coupling $\xi|\Phi|^2R$ to gravity, both with and without the electroweak gauge fields coupled to the Higgs. We assume a minimal scenario in which inflation and reheating are caused by chaotic inflation with a quadratic potential, and no additional new physics is relevant below the Planck scale. By using classical real-time lattice simulations with a renormalisation group improved effective Higgs potential and by demanding the stability of the Higgs vacuum after inflation, we obtain upper bounds for $\xi$, taking into account the experimental uncertainty of the top-Yukawa coupling. We compare the bounds in the absence and presence of the electroweak gauge bosons, and conclude that the addition of gauge interactions has a rather minimal impact. In the unstable cases, we parametrize the time when such instability develops. For a top-quark mass $m_t \approx173.3 {\rm GeV}$, the Higgs vacuum instability is triggered for $\xi \gtrsim 4 -5$, although a slightly lower mass of $m_t \approx 172.1 {\rm GeV}$ pushes up this limit to $\xi \gtrsim 11 - 12$. This, together with the estimation $\xi \gtrsim 0.06$ for stability during inflation, provides tight constraints to the Higgs-curvature coupling within the SM.
\end{abstract}

\maketitle

\section{Introduction}
\label{sec:i}

The Standard Model (SM) potential may become negative at very high energies \cite{Bezrukov2012, Degrassi2012}. This has prompted an important effort to determine whether the electroweak vacuum is, in the present, stable or unstable. Current measurements of the top quark and Higgs masses suggest that we live in a metastable Universe: the probability of the Higgs field to decay into a higher-scale negative-energy vacuum is non-zero, but the estimated decay time is much larger than the present age of the Universe~\cite{Espinosa:2007qp}.

However, the situation is quite different in the early Universe. In this case, high energies and high spacetime curvature can make the vacuum more unstable. In particular, this may happen during inflation~\cite{Espinosa:2007qp,Kobakhidze:2013tn,Fairbairn:2014zia,Enqvist:2014bua,Herranen2014,Kearney:2015vba,Kamada:2014ufa,Espinosa:2015qea,Joti:2017fwe,Rajantie:2017ajw}, or during the successive period of (p)reheating~\cite{Herranen2015,Kohri2016,Ema2016, Enqvistetal2016, Ema:2017loe}. The dynamics of the Higgs field $\Phi$ during and after inflation, as well as the potential instability of the Higgs vacuum, depend very sensitively on the strength of its non-minimal coupling to the scalar curvature, defined as $\xi|\Phi|^2R$, with $R$ the Ricci scalar. This interaction is necessary to renormalise the theory in curved space~\cite{Parker2009,Birrell1984}, and given that $\xi$ runs with energy, it cannot be set to zero at all energy scales. Gravitation is very weak in comparison with the other interactions, so current particle-physics experiments provide only very weak constraints to this coupling \cite{Atkins2013}. The coupling $\xi$ can be considered, therefore, as the last unknown parameter of the SM.

If $\xi \lesssim 0.1$ the Higgs is effectively light during inflation, and behaves as a spectator field, forming a condensate with a large
vacuum expectation value (VEV)~\cite{Starobinsky1994,Espinosa:2007qp,DeSimone:2012qr,Enqvist2013}. If it exceeds the position of the potential barrier, the Higgs reaches its true negative-energy vacuum and generates patches of anti-de Sitter space, resulting in a catastrophic outcome for our Universe~\cite{Espinosa:2007qp,Kobakhidze:2013tn,Fairbairn:2014zia,Enqvist:2014bua,Herranen2014,Kearney:2015vba,Espinosa:2015qea,Joti:2017fwe,Rajantie:2017ajw}. 
One simple way to prevent this from happening is to consider a sufficiently low inflationary scale, so that even if the Higgs is excited during inflation, its amplitude never reaches the potential barrier. Another way of ensuring vacuum stability is to consider values of the top quark mass 2-3 sigma below its central value $m_t = 172.44^{\pm 0.13 {\rm (stat)}}_{\pm 0.47 {\rm (syst)}} {\rm GeV}$ \cite{Khachatryan:2015hba}, so that the second minimum in the Higgs potential is either shifted to sufficiently high energies, or it is simply not present. In any case, if the Higgs field remains stable during inflation, it starts oscillating around the minimum of its potential shortly after inflation ends, rapidly decaying into the SM gauge bosons and fermions via non-perturbative parametric effects~\cite{Enqvist2013,Figueroa:2014aya,Figueroa2015,Enqvist2015}. This may have relevant cosmological consequences, like the realization of successful reheating into the SM without additional mediator fields~\cite{Figueroa:2016dsc}, the realization of baryogenesis via leptogenesis in certain extensions of the SM~\cite{Kusenko:2014lra,Yang:2015ida,Pearce:2015nga}, or the production of a primordial background of gravitational waves (GW) peaked at high frequencies~\cite{Figueroa:2014aya,Figueroa:2016ojl}. 

On the other hand, if $\xi \gg 0.1$, the height of the potential barrier increases, and the Higgs is no longer a light degree of freedom during inflation~\cite{Espinosa:2007qp,Herranen2014}. In this case, the Higgs field acquires an effective mass of the order of $m_{\Phi}^2 \simeq \xi R \sim 12 \xi H_*^2 \gtrsim H_*^2$ during inflation, with $H_*$ the inflationary Hubble rate. This prevents the Higgs from developing large amplitude fluctuations during inflation. However, the situation is quite the opposite after inflation ends. The post-inflationary oscillations of the inflaton $\phi$ around the minimum of its potential induce rapid changes in the spacetime curvature 
$R$, which becomes negative during a significant fraction of time in each oscillation. The effective mass of the Higgs field becomes tachyonic during those moments, $m_{\Phi}^2 \propto R < 0$. If $\xi$ is sufficiently large, the Higgs field may be significantly excited during the tachyonic periods, potentially triggering the vacuum instability~\cite{Herranen2015}. This issue has been studied lately, using both analytical and numerical techniques, as well as classical real-time lattice simulations \cite{Herranen2015,Kohri2016,Ema2016}. The results of all of these works agree qualitatively, finding $\xi \lesssim \mathcal{O} (1) - \mathcal{O} (10)$ as an upper bound for achieving stability after inflation. A similar lattice analysis of the values of the Higgs-inflaton coupling inducing the instability of the Higgs vacuum has also been carried out in \cite{Enqvistetal2016}, while an analysis of the combined effects of both Higgs-curvature and Higgs-inflaton couplings has been done in~\cite{Ema:2017loe}. 

In this work, we use classical field theory lattice simulations to constrain the range of allowed $\xi$ values which ensure the stability of the Higgs vacuum after inflation. We do a systematic parameter analysis of the Higgs post-inflationary dynamics. We use in the simulations the renormalization group improved Higgs effective potential, and study the impact of the initial conditions and number of Higgs components in the results. We include also an analysis of how the time scale at which the Higgs field develops the instability depends on $\xi$ and the top-quark mass.  

Furthermore, we consider the more realistic situation, so far not analyzed in the literature, where the Higgs field is coupled to the electroweak gauge bosons. We mimic the SM gauge interactions with an Abelian-Higgs analogue model, but argue that this is enough to demonstrate the effect of restoring the gauge interactions. Our Abelian set-up captures well the gauge boson field effects onto the Higgs post-inflationary dynamics, as we expect the non-Abelian terms of the Lagrangian to be subdominant, especially at the earliest times. In this paper we assess for the first time the implications for the $\xi$ bounds due to the presence of the SM electroweak interactions.

We have assumed throughout this work a chaotic inflationary model ${\frac{1}{2}}m_{\phi}^2 \phi^2$, with $m_{\phi} \simeq 10^{13}$ GeV fixed by the observed amplitude of the CMB anisotropies~\cite{Ade:2015lrj}. We note that, although the simple chaotic inflation model with a quadratic potential is in tension with observational data~\cite{Ade:2015lrj,Tsujikawa:2013ila}, it is possible to modify the large field behavior to make it consistent with observations~\cite{Tsujikawa:2013ila,Saha:2016ozn}, without changing its post-inflationary dynamics.

The structure of the paper is as follows. In Section \ref{sec:ii} we present a brief review of the inflaton and Higgs dynamics after inflation in the presence of a Higgs-curvature non-minimal coupling. In Section \ref{sec:iii} we present the equations of motion and the initial conditions of the different fields, as well as  some qualitative aspects of our lattice simulations. The following three sections present the results from our lattice simulations, with increasing degree of complexity. In Section \ref{sec:iv} we consider a free scalar field with no potential. This is useful to understand better the results in Section \ref{sec:v}, where we introduce the renormalisation group improved Higgs potential. We determine the values of the coupling $\xi$ that give rise to an unstable Universe, and parametrize the time scale at which the instability takes place, as a function of $\xi$ and $m_t$. In Section \ref{sec:vi} we repeat the same analysis, but including also the gauge bosons in the lattice. In Section \ref{sec:vii} we discuss our results and conclude. Finally, in the Appendix we discuss the details of the lattice formulation we have used in our simulations.

In this paper we use the signature $(-,+,+,+)$ and, hence, consider a flat background with Friedman-Lem\^aitre-Robertson-Walker (FLRW) metric like {\small$ds^2 =  -dt^2 + a^2 (t) dx^i dx^i$}, where {\small$a(t)$} is the scale factor, and {\small$t$} is the cosmic time. We refer to the reduced Planck mass as $m_p = (8 \pi G)^{-1/2} \simeq 2.44\cdot 10^{18}$ GeV.  

\section{Higgs excitation due to Inflaton oscillations}\label{sec:ii}

\begin{figure}
      \begin{center} 
      \includegraphics[width=8cm]{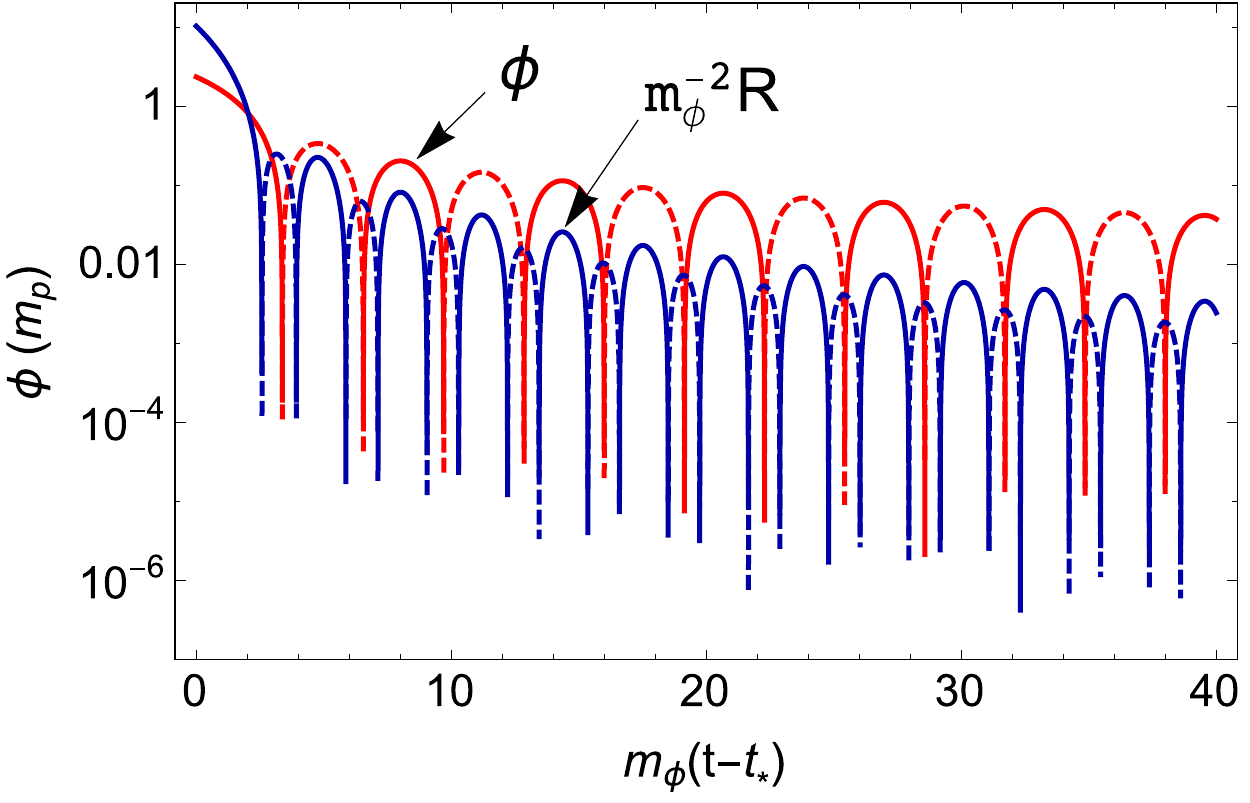}\vspace{0.2cm}
      \end{center} 
\caption{ The red line shows the oscillations of the inflaton field as a function of time  in units of $m_p = (8 \pi G)^{-1/2}$, and the blue line shows, for comparison, the corresponding (dimensionless) Ricci scalar $m_{\phi}^{-2} R $. A solid line indicates positive values, and a dashed line negative values.} 
            \label{fig:inflaton}
 \end{figure}

We consider throughout the paper the inflationary chaotic model $V(\phi) = \frac{1}{2} m_{\phi}^2 \phi^2$, where $\phi$ is the inflaton, and its mass is fixed to approximately $m_{\phi} = 1.5 \times 10^{13} {\rm GeV}$ in order to explain the CMB anisotropies. If $\phi \gtrsim \mathcal{O}(10)m_p$, the field is in a slow-roll regime, causing the inflationary expansion of the Universe. However, when  $H(t) \approx m_{\phi}$ with $H(t)$ the Hubble parameter, the inflaton field starts oscillating around the minimum of its potential, ending the inflationary stage. Let us define $t_*$ as the time when $H(t_*) = m_{\phi}$ holds exactly, and consider this moment as the onset of the inflaton oscillations. The coupled equations of motion of the inflaton and scale factor are
\bea \ddot{\phi} + 3 H(t) \dot{\phi}  + m_{\phi}^2 \phi = 0 \ ,\label{eq:eom-infl}\\
H^2 (t) \equiv \left({\dot{a}\over a}\right)^2 = \frac{1}{6 m_p^2}(\dot{\phi}^2 + m_{\phi}^2 \phi^2 )  \ .\label{eq:eom-Hub}
\eea 
To obtain the initial conditions for the homogeneous inflaton, we have solved numerically the coupled inflaton and Friedmann equations, Eq.~(\ref{eq:eom-infl}) and Eq.~(\ref{eq:eom-Hub}), imposing the slow-roll conditions $\dot\phi \simeq - m_\phi^2\phi^2/3H^2$, $\dot\phi \ll m^2_\phi\phi^2$ well before the end of inflation. From the numerical solution, we obtain the time $t_*$ when $H(t_*) = m_{\phi}$ holds exactly. At this moment we find
\be \phi (t_*) \simeq 2.32 m_p \ , \hspace{0.3cm} \dot{\phi} (t_*) \simeq  -0.78 m_{\phi} m_p \ .\ee 
Using Eqs.~(\ref{eq:eom-infl}), (\ref{eq:eom-Hub}), the Ricci scalar can be expressed  in terms of $\phi$ and $\dot\phi$, like
\be R(t) \equiv 6 \left[ \left( \frac{\dot{a}}{a} \right)^2 + \frac{\ddot{a}}{a} \right] = \frac{1}{m_p^2} ( 2 m_{\phi}^2 \phi^2 - \dot{\phi}^2 ) \ . \label{eq:ricci-eom} \ee
The inflaton field after inflation behaves, approximately, as a damped oscillator with a decaying amplitude \cite{Kofman1997}:
\begin{eqnarray}
\phi (t) \simeq \phi_a (t) \sin(m_{\phi} t) \ , \hspace{0.5cm} \phi_a (t) = \sqrt{\frac{8}{3}} \frac{m_{p}}{m_{\phi}t} \label{eq:PHI-term} \ .
\end{eqnarray}
Each time the inflaton field crosses around zero, $\phi \approx 0$, we have $R(t) < 0$ from Eq.~(\ref{eq:ricci-eom}). This can be clearly seen in Fig.~\ref{fig:inflaton}, where we plot both the inflaton and the Ricci scalar as a function of time. 

Let us focus now on the post-inflationary dynamics of the Higgs field. The relevant piece of the Standard Model we need to consider is
\be \mathcal{L}_{\rm SM} \supset \int d^4 x a^3 \left( \frac{1}{4 g_s^2} F_{\mu \nu} F^{\mu \nu} + |D_{\mu} \Phi |^2  + \xi R |\Phi |^2 + V(\Phi) \right) \label{eq:action-SM} \ee
where $V(\Phi) \equiv \lambda\left(\Phi^\dag\Phi-v^2/2\right)^2$ is the Higgs potential, $v = 246 {\rm GeV}$ is the electroweak vacuum expectation value (VEV), $\xi$ is the strength of the coupling of the Higgs to the Ricci scalar, $F_{\mu\nu}$ the gauge field strength (assumed Abelian for simplicity), and $D_{\mu} = \partial_\mu - i A_\mu /2 $ the gauge covariant derivative describing the interaction of the Higgs field with the electroweak gauge bosons. Because of the coupling to the scalar curvature, the Higgs gets an effective mass $m_{\Phi}^2 (t) = \xi R(t)$. Therefore, the Higgs becomes effectively tachyonic with $m_{\Phi}^2 < 0$, during the intervals when the Ricci scalar becomes negative. Because of this, there is a strong periodic excitation of the Higgs field, a phenomenon known as tachyonic resonance~\cite{Dufaux2006}. 

We can estimate both the period of time that the Ricci scalar becomes negative, as well as the maximum momenta excited by the resonance. The inflaton crosses zero periodically at $m_\phi t_n = (n-1/2)\pi$, $n = 1, 2, 3, ...$. We can determine a typical envelope amplitude between the $n$-th and the $(n+1)$-th crossings, as $\phi_n/m_p = {\sqrt{8/3}/\pi n}$. When the inflaton crosses around zero, the Ricci scalar becomes negative $R \simeq \phi_n^2(m_\phi/m_p)^2(3m_\phi^2\Delta t^2 - 1) < 0$ for a time $m_\phi\Delta t \lesssim {2/\sqrt{3}} \approx 1.2$, while the inflaton amplitude is $|\phi| \lesssim \phi_n /\sqrt{3} \sim 0.3 m_p / n$.  On the other hand, the greater the coupling $\xi$, the larger the range of Higgs tachyonic modes excited while the curvature is negative. We estimate this as an infrared (IR) band from $k = 0$ up to a cutoff $\Lambda$, $k \in [0, \Lambda]$, with
\be \Lambda \simeq {2\sqrt{2}\over\sqrt{3}}{a_n\over\pi n}\sqrt{\xi}\ , \label{eq:momenta-cutoff}\ee
where $a_n$ is the scale factor at $t_n$ (we take initially $a_1 = 1$). Let us consider the unitary gauge, so that the SM Higgs doublet can be written as a real degree of freedom, $\Phi = \varphi/\sqrt{2}$. Let us redefine the Higgs amplitude as $h \equiv \varphi / a^{3/2}$ so that in cosmic time, this rescaling eliminates the friction term in the Higgs equation of motion (EOM). If we ignore the presence of the gauge bosons and of the Higgs self-interacting potential, the equation of motion of its Fourier modes is
\be \ddot{h}_k + \left[ \frac{k^2}{a^2} + \xi R (t) + \Delta \right] h_k = 0\,,~~h_k \equiv {\varphi_k\over\sqrt{2}} \label{eq:higgs-mode-eq}\ee
where $\Delta \equiv -\frac{3}{4} \frac{\dot{a}^2}{a^2} - \frac{3}{2} \frac{\ddot{a}}{a}$, so that $\Delta \ll k^2/a^2$ for sub-horizon scales. We can then set $\Delta \rightarrow 0$, and using Eqs.~(\ref{eq:ricci-eom})-(\ref{eq:PHI-term}), write the previous EOM as a Mathieu equation
\be \frac{d^2 h_k}{d z^2} + (A_k - 2q \cos(2 z) ) h_k = 0 \ , \label{eq:higgs-Mathieu}\ee
where $z \equiv m_{\phi} (t - t_*)$ and
\be A_k \equiv \frac{k^2}{a^2 m_{\phi}^2 } + \frac{\phi_n^2 (z)}{2 m_{p}^2} \xi \ , \hspace{0.3cm} q \equiv \frac{3 \phi_n^2 (z)}{4 m_{p}^2} \left( \xi - \frac{1}{4} \right) \ . \label{eq:q-term}\ee
These equations of motion have been extensively studied in the context of parametric resonance in $m_{\phi}^2 \phi^2$ preheating, see for example \cite{Kofman1994, Kofman1997}. The main difference with respect to standard parametric resonance is that we are not constrained now to the case $A_k > 2q$, and hence we have greater resonance bands which induce a stronger particle creation effect in the broad resonance regime $q \gg 1$. However, note that due to the expansion of the Universe, $\phi_n (z)$ decreases, and hence this pushes the Higgs into a narrow resonance regime, where this effect is much weaker. The dynamics of this theory was studied in~\cite{Herranen2014} with the properties of tachyonic resonance of \cite{Dufaux2006}, and after that numerically in~\cite{Kohri2016} and in the lattice in \cite{Ema2016,Enqvistetal2016,Ema:2017loe}. 

\subsection{Higgs potential}

\begin{figure}
      \begin{center} 
      \includegraphics[width=8.2cm]{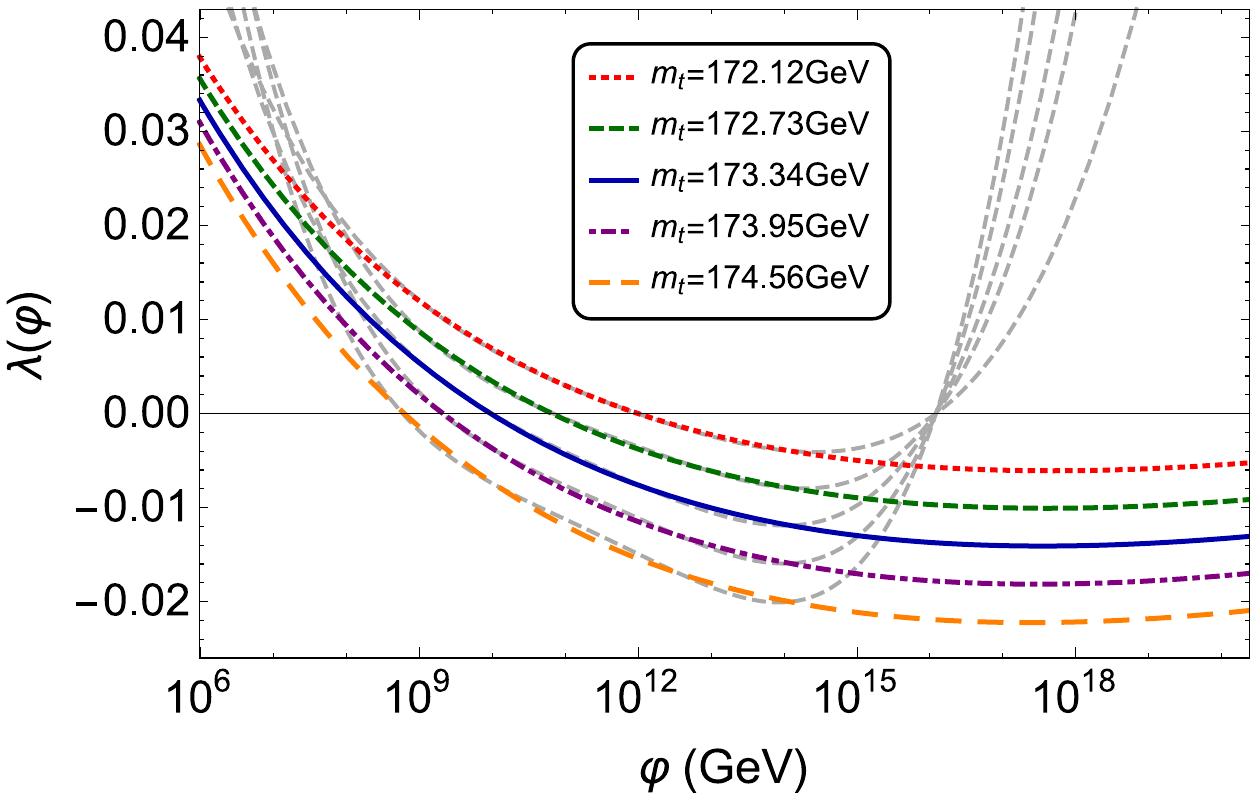}
      \end{center} 
      \caption{Running of $\lambda(\varphi)$ as a function of the Higgs field $\varphi$ for $\alpha_s=0.1184$, $m_h=125.5 {\rm GeV}$, and different values of the top quark mass $m_t$, obtained from the public package in \cite{Bezrukov}. The corresponding gray dashed lines indicate the interpolation $\lambda_{\rm in} (|\varphi | )$ used in the lattice simulations.} 
            \label{fig:lambda-run}
 \end{figure}

Let us consider now the effect of plugging back the Higgs potential. In particular we consider the renormalisation group improved Higgs potential
\be V(\varphi) = \frac{\lambda (\varphi)}{4} \varphi^4 \ ,  \label{eq:higgs-potential}\ee
valid for large field amplitudes above the electroweak scale, $\varphi \gg v \equiv 246~{\rm GeV}$. 
Here $\lambda (\varphi)$ is the renormalised Higgs self-coupling at the renormalisation scale $\mu=\varphi$, where the running behavior has been computed up to three loops \cite{Bezrukov2012, Degrassi2012}. 
The running is very sensitive to the strong coupling constant $\alpha_s$, Higgs mass $m_h$, and Yukawa top coupling $y_t$, the latter being currently the strongest source of uncertainty. We show in Fig.~\ref{fig:lambda-run} the running of $\lambda (\varphi)$ for the central values $\alpha_s=0.1184$, $m_h=125.5 {\rm GeV}$, and different values of the Yukawa top quark coupling. The figure has been obtained with the public package of \cite{Bezrukov}. We observe that the Higgs potential possesses a maximum (a barrier) at a given scale $\varphi_+$, and crosses zero at $\varphi_{\rm o}$, becoming negative at higher scales. We show these values for different top quark masses in Table \ref{table:higgspot}. For the world-average top quark mass $m_t = 172.44^{\pm 0.13 {\rm (stat)}}_{\pm 0.47 {\rm (syst)}} {\rm GeV}$~\cite{Khachatryan:2015hba}, 
we have $\varphi_{\rm o} \approx  10^{11} {\rm GeV}$. Moving this mass $\sim 1.9$ sigma below, we have $\varphi_{\rm o} \approx  10^{14} {\rm GeV}$, while for $\sim 2.5$ sigma below $\varphi_{\rm o}$ is pushed to infinity. Let us also note that the effective potential also depends on the spacetime curvature through loop corrections, but as seen in \cite{Markkanen:2018bfx}, these terms are relevant only for small couplings $\xi \lesssim 1$.

Let us now incorporate the potential into the Higgs mode equation,
\be \ddot{h}_k + \left[ \frac{k^2}{a^2} + \xi R (t) + \Delta + \frac{\lambda (\varphi)}{a^3} \langle h^2 \rangle\right] h_k = 0 \ . \label{eq:higgs-mode-eq2} \ee
If $\lambda >0$, the Higgs tachyonic resonance effect weakens, as the Higgs self-interaction $\lambda(\varphi)\langle h^2 \rangle > 0$ compensates the negativeness of $\xi R < 0$. If $\lambda <0$, the tachyonic effect, on the contrary, is enhanced. The presence of the Higgs potential represents a correction over the mode excitation described by Eqs.~(\ref{eq:higgs-mode-eq}), (\ref{eq:q-term}). We need therefore to introduce the system into a lattice, where we can solve numerically the EOM of the Higgs including its own potential non-linearities, taking into account both cases $ \lambda< 0$ and $\lambda > 0$, changing locally due to the running of $\lambda$, depending on the Higgs value at every lattice site. Before we move on into the simulation details, we need to add a step further, restoring the electroweak gauge interactions of the Higgs.

\begin{table}  
\begin{center}
  \begin{tabular}{ | c | c | c |}
    \hline
    $m_t ({\rm GeV}) $ & $\varphi_{+} ({\rm GeV})$ & $\varphi_{0} ({\rm GeV})$  \\ \hline
    172.12 & $7.83 \times 10^{11}$ & $1.01 \times  10^{12}$  \\  \hline
    172.73 & $5.20 \times 10^{10}$ & $6.70 \times  10^{10}$  \\  \hline
    173.34 & $7.49 \times 10^{9}$ & $9.65 \times  10^{9} $  \\  \hline
    173.95 & $1.67 \times 10^{9}$ & $2.15 \times  10^{9} $  \\  \hline
   174.56 & $4.92 \times 10^{8}$ & $6.34 \times  10^{8} $  \\  \hline
  \end{tabular}
\end{center}
\caption{The values of the Higgs field where the potential Eq.~(\ref{eq:higgs-potential}) has a maximum ($\varphi_+$) and crosses zero ($\varphi_0$), obtained for $\alpha_s=0.1184$, $m_h=125.5 {\rm GeV}$, and different values of the top quark mass. These quantities have been obtained with the public package of \cite{Bezrukov}.} \label{table:higgspot}
 \end{table}

\subsection{Electroweak gauge interactions}
\label{sec:iii}


The equations of motion in the continuum can be derived from the minimization of action (\ref{eq:action-SM}). We take the Higgs $\Phi = \varphi/\sqrt{2}$ as a complex doublet, with four real components
\begin{equation}
     \varphi =\begin{pmatrix}
         \varphi_1 + i \varphi_2 \\
         \varphi_3 + i \varphi_4
        \end{pmatrix} \ , \hspace{0.2cm} \varphi_n \in \mathfrak{Re} \ .
  \end{equation}
As we will argue later, we have neglected the purely non-Abelian terms of the gauge field self-interactions. The covariant derivative is then simply written as $(D_{\mu})_{ab} \equiv \delta_{ab} (\partial_{\mu} - i A_{\mu} )$, with $a,b=1,2$. The equations of motion, in the temporal gauge $A_0=0$, are 
\bea \label{eq:eom-fulleom1}
&& \ddot{\Phi} - \frac{1}{a^2} D_i D_i \Phi + 3 \frac{\dot a}{a} \dot{\Phi} + 2  [ \xi R +  \lambda (| \varphi | ) (\Phi^{\dagger} \Phi ) ]\Phi  \nonumber \\
\label{eq:eom-fulleom2}
&& \hspace{4cm} + \frac{\partial \lambda}{\partial | \varphi |} (\Phi^{\dagger} \Phi)^2 = 0 \ , \\
\label{eq:eom-fulleom3}
&& \ddot{A}_j - \frac{1}{a^2} ( \partial_i \partial_i A_j - \partial_i \partial_j A_i ) + \frac{\dot a}{a} \dot{A}_j = 2 g_s^2 \IM [\Phi^{\dagger} (D_j \Phi )] \ , \nonumber \\\\
&& \hspace{4cm} \partial_i \dot{A}_i = 2 g_s^2 a^2 \IM[\Phi^{\dagger} \dot{\Phi} ]  \ . \label{eq:eom-fulleom} \eea
The third of these equations is the Gauss constraint, a relation between fields that must be obeyed at all times. In the lattice we solve a discrete version of equations~(\ref{eq:eom-fulleom1})-(\ref{eq:eom-fulleom}), obtained from a discrete gauge-invariant action, see Eqs.~(\ref{eq:eom-gaugediscrete}), (\ref{eq:gauss-discrete}). Details of the lattice formalism are given in the Appendix. Note also that this is not, strictly speaking, the standard Abelian-Higgs model, as we are introducing two Higgs complex fields instead of just one. 

The form of $a(t)$ in these equations, as well as the Ricci scalar $R(t) = R[a, \dot{a}, \ddot{a} ]$, is obtained from the self-consistent solution of the inflaton and Friedmann equations (\ref{eq:eom-infl}) and (\ref{eq:eom-Hub}). As we shall see, for the values of $\xi$ considered in this work, the energy of the Higgs field is always several orders of magnitude subdominant with respect to the energy of the inflaton. Hence, we just ignore the contribution of the Higgs field to the Friedmann equation. Note that the inflaton is taken as a homogeneous field, and we do not introduce it explicitly in the lattice, it simply dictates the form of $a(t)$ and $R(t)$ as a function of time.

In Section \ref{sec:iv} we study tachyonic resonance in the lattice, taking the Higgs as a free field without self-interaction. The Higgs will then be excited only due to the rapidly changing spacetime background. In Section \ref{sec:v} we re-introduce back the Higgs potential, but ignore yet its interaction with the gauge bosons. We determine under those circumstances, what values of $\xi$ lead the Higgs field to become unstable, so that it rolls rapidly into the true vacuum. In Section \ref{sec:vi} we finally incorporate a gauge structure into the simulations, and study their effect on the post-inflationary Higgs dynamics, re-evaluating again the critical values of $\xi$. 

Choosing $\mu = |\varphi|$, we obtain the renormalisation group improved effective potential\footnote{As argued in Ref.~\cite{Herranen2014}, the scale choice should also involve the Ricci scalar $R$, but in the current time-dependent case it could lead to unphysical effects.}, introducing in the lattice the Higgs potential evaluated as 
\be V( | \varphi | ) = \frac{\lambda(| \varphi | )}{4} |\varphi |^4 \ , \hspace{0.5cm} | \varphi | = \sqrt{\sum_{n=1}^4 \varphi_n^2 } \ ,\ee
where we assume $|\varphi| \gg v$. As seen before, the Higgs self-coupling $\lambda (\varphi)$ runs with the value of $\varphi$. We introduce the running in our simulations as a local function of the lattice point $n$, i.e.~$\lambda(|\varphi(n)|)$: as the value of $|\varphi|$ changes from lattice point to lattice point, so does too the value of the Higgs self-interaction. More specifically, we introduce a quartic logarithmic polynomial $\lambda_{\rm in} (|\varphi|) = \sum_{n=0}^4 c_n \left( \log |\varphi | \right)^n$, interpolating the 3-loop calculation of the running obtained in \cite{Bezrukov} for the relevant range of Higgs amplitudes $|\varphi|$ (see Fig.~\ref{fig:lambda-run}).  As we have mentioned, the running of the potential depends strongly on the value of the top quark mass, the current world average being $m_t= 172.44^{\pm 0.13 {\rm (stat)}}_{\pm 0.47 {\rm (syst)}}~{\rm GeV}$~\cite{Khachatryan:2015hba}. We take this uncertainty into account by providing different sets of $\lbrace c_n \rbrace$ constants, corresponding to different interpolations of the running for each value of $m_t$. 

Our interpolation can describe appropriately only the running of $\lambda$ for certain values of $|\varphi|$, failing at low and high field amplitudes. This is, however, not a problem, because those field values are never reached anywhere in the lattice, before the instability of the Higgs field is developed. On the other hand, when the Higgs has become unstable and decays towards the negative-energy vacuum, the amplitude of the Higgs field starts increasing very fast, reaching the region where the interpolation fails. However, our aim in this work is to determine the specific time when the instability is developed, not to characterize the dynamics of the Higgs field once the instability has commenced. In fact, in order to ensure numerical stability during the Higgs field transition from positive to negative $\lambda$, it is convenient to modify the high-energy running of the latter, so that it generates a second vacuum at an energy lower than that dictated by the real running predicted in the Standard Model. This is achieved for $c_4 > 0$. In particular, we have chosen the constants so that the negative-energy vacuum is generated at approximately $\varphi = \varphi_v \approx 10^{16} {\rm GeV}$. If the Higgs amplitude goes to this vacuum with negative potential energy, we say that the Higgs has become unstable. We have explicitly checked that our characterization of the times of instability is independent on the particular choice of constants $c_n$ (for a given $m_t$ value), as long as they fit the Higgs effective potential within the range $\sim 10^9-10^{14}$ GeV.

\subsection{Initial conditions}

We start the lattice simulations at time $t = t_*$, where we impose for all four components of the Higgs that their initial homogeneous amplitude vanishes: $ \varphi_n (t_*) = 0$, $n=1,2,3,4$. We then add on top a spectrum of fluctuations, which mimic the spectra of quantum vacuum fluctuations\footnote{Our initial conditions are set at a time when the slow-roll conditions are not yet totally broken. Therefore, we can introduce instead quantum vacuum fluctuations in {\it de~Sitter},
\be \langle | \varphi_k |^2 \rangle = \frac{\pi e^{-\pi \IM [\nu] }}{4 H_* a_*^3}  \left| H_{\nu}^{(1)} \left(\frac{k}{a_* H_*} \right) \right|^2  \ \ee
with $\nu = \sqrt{9/4 - ( \xi R_* / H_*)^2 } $. However, for the couplings $\xi  > 4$ we are considering, this spectra is almost identical to the FLRW case described by Eq.~(\ref{eq:FRWquanVacFluc}).},
\be \langle | \varphi_k |^2 \rangle = \frac{1}{2 a_*^3 \omega_k}  \ , \hspace{0.5cm} \omega_k = \sqrt{\frac{k^2}{a_*^2} +  \xi R_*} \ ,\label{eq:FRWquanVacFluc}\ee
where $a_* = a(t_*) \equiv 1$, and $R_* \equiv R(t_*) \approx 10 H_*^2$ from Eq.~(\ref{eq:ricci-eom}).

The spectra of quantum fluctuations (\ref{eq:FRWquanVacFluc}) is set in the lattice in a similar way as in \emph{Latticeeasy}~\cite{Felder2008}, imposing in momentum space the following spectra for the Higgs field amplitude and derivatives 
\bea \varphi_n (k) &=& \frac{|\varphi_n|}{\sqrt{2}} (e^{i \theta_{n1} } + e^{i \theta_{n2}} ) \ , \hspace{0.3cm} (k< k_c)  \nonumber \\ 
\varphi'_n (k) &=& \frac{|\varphi_n|}{\sqrt{2}} i \omega_{k,n} (e^{i \theta_{n1} } - e^{i \theta_{n2}}  ) \ , \hspace{0.3cm} (k< k_c) \label{eq:init-spectra} \eea
where $\omega_{k,n} \equiv \sqrt{(k/a_*)^2 + \xi R_*}$, $\theta_{n1}$ and $\theta_{n2}$ are real phases drawn from a uniform random distribution in the interval $\theta_{n1},\theta_{n2} \in [0 , 2 \pi)$, whereas $|\varphi_n|$ varies according to the probability distribution
\be P(|\varphi_n| ) d |\varphi_n| = \frac{2 |\varphi_n|}{\omega_{k,n}^2} e^{- \frac{|\varphi_n|^2}{\omega_{k,n}^2} } d |\varphi_n| \ . \label{eq:probdist} \ee
The ultraviolet cutoff $k_c$ is introduced  in order to prevent the excitation of UV modes which are not expected to be excited by the tachyonic resonance, i.e. $k_c\approx\Lambda$ with $\Lambda$ given by Eq.~(\ref{eq:momenta-cutoff}). 

Hence, the variance of (a component of) the Higgs field initially is 
 \bea \langle \varphi_*^2 \rangle &=& \frac{1}{4 \pi^2 a_*^3} \int_0^{k_c} dk \frac{k^2}{\omega_k} \nonumber \\
 &=& \frac{1}{8 \pi^2} \left( k_c \omega_{k_c} + \xi R_* \log \left[ \frac{\xi R_*}{k_c + \omega_{k_c}} \right] \right) \ ,\label{eq:quant-fluct} \eea
where we have taken $a_*=1$ in the second equality. Typical numbers chosen in our simulations are $\xi \sim 10$ and $k_c \sim 10 H_*$, which gives an initial Higgs amplitude of
\be \sqrt{\langle \varphi_*^2 \rangle } \approx 0.82 H_{*} \approx 1.2 \times 10^{13} \rm{GeV}\label{eq:quant-fluct2} \ . \ee
Typically $\sqrt{\langle \varphi_*^2 \rangle } \gg \varphi_+$, and hence most or the entire Higgs field is already in the right side of the barrier when initial conditions are set. This, however, does not mean that the Higgs field will immediately become unstable, as the mainly positive sign of $R$ may impede it. We shall discuss this issue in more detail in Sections \ref{sec:v} and \ref{sec:vi}. Let us also remark that this way of fixing the initial conditions is appropriate only if the tachyonic resonance regime of the system enhances the Higgs amplitude significantly over the value given in Eq.~(\ref{eq:quant-fluct2}). If it does not, we cannot trust the lattice approach. Finally, let us also note that there is a contribution to the Higgs effective mass from its self-interactions, i.e.~the effective Higgs mass should be rather $m_{\rm eff}^2 \approx \xi R_* + \lambda \langle \varphi_* \rangle^2$. Taking $\lambda \approx - 0.01$, $\xi \approx 10$, and $H_* = m_{\phi} \approx 6 \times 10^{-6} m_p$, we see that the second term (Higgs self-interaction) is negligible with respect to the first one (Higgs non-minimal coupling).

\section{Simulations with a free scalar field}\label{sec:iv}

\begin{figure*}
      \begin{center} 
      \includegraphics[width=7.6cm]{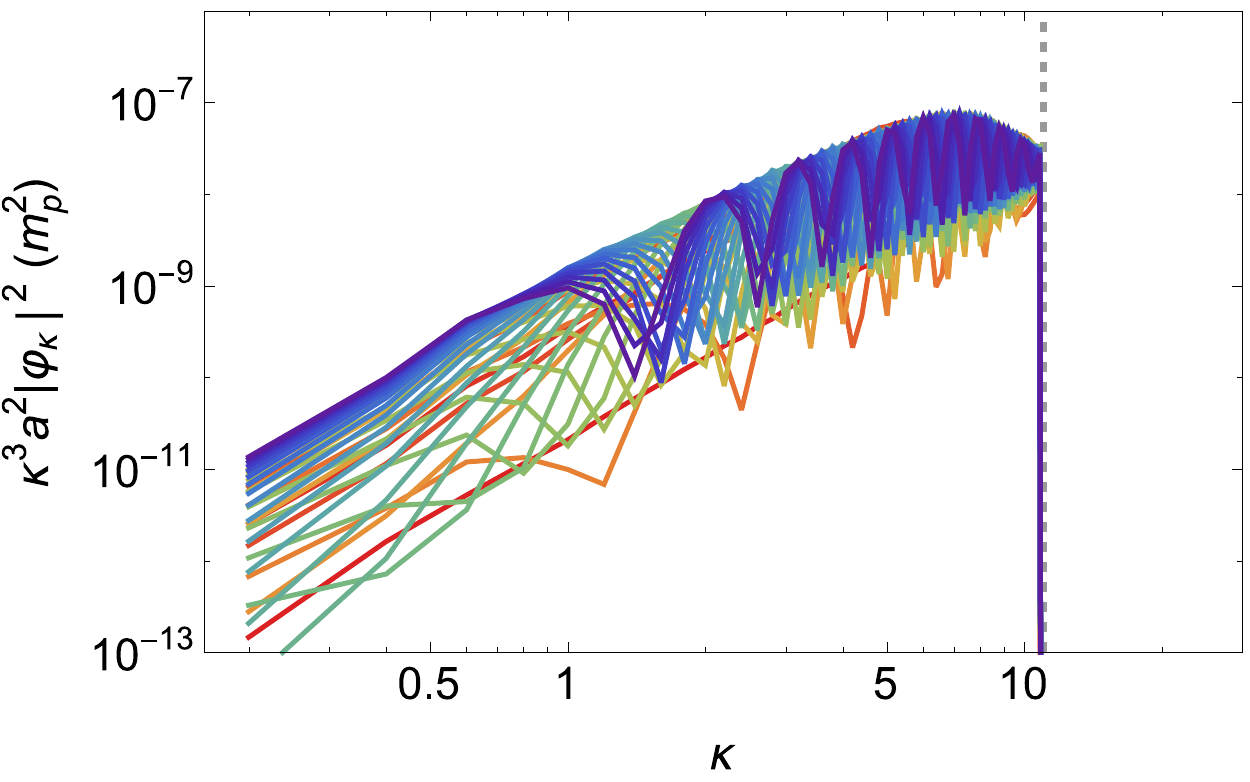}\hspace{0.6cm}
       \includegraphics[width=7.6cm]{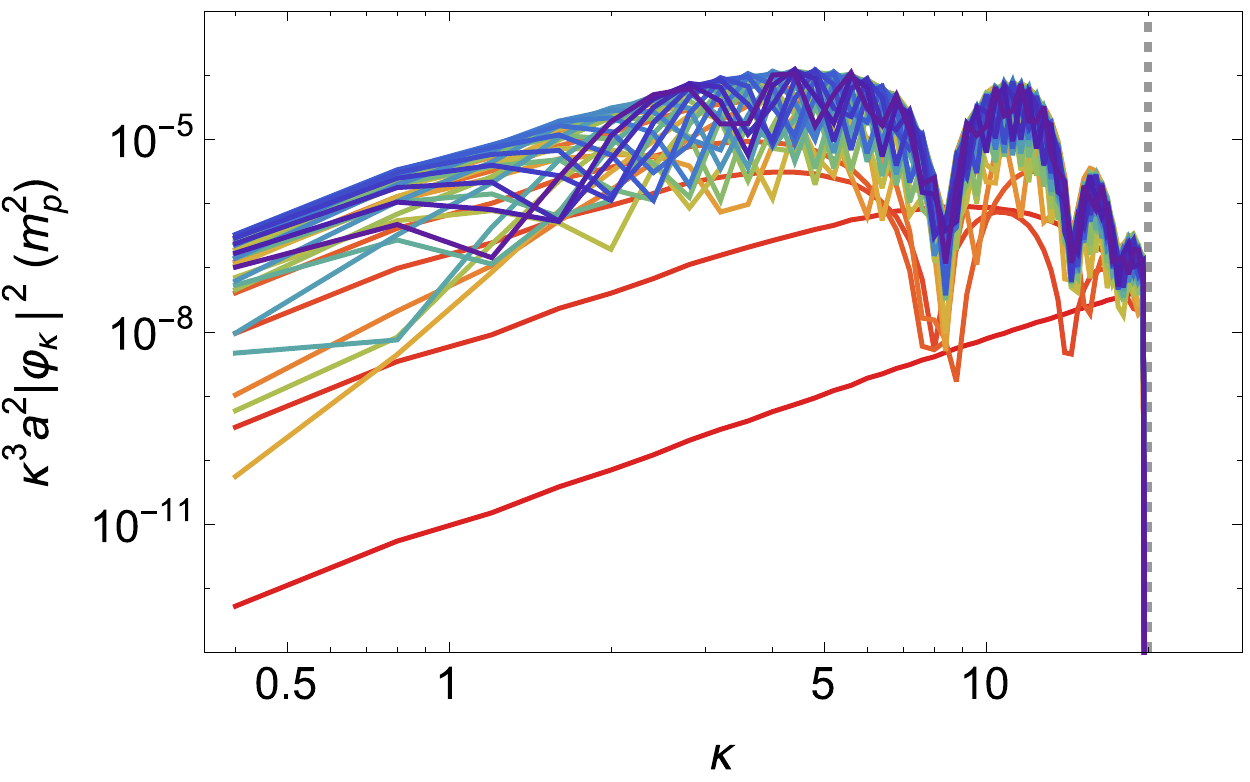}
      \end{center} 
      \caption{The time evolution of the Higgs field spectra $\kappa^3 a^2 |\varphi_{\kappa}|^2$ as a function of $\kappa \equiv k/m_{\phi}$, for the non-interacting case (Section~\ref{sec:iv}) with the Higgs-curvature couplings $\xi=5$ (left panel) and $30$ (right panel). The different coloured lines show the spectra at different times, going from early times (red) to late times (purple). The time interval between lines is $m_{\phi} \Delta t = 2$, so $m_{\phi} (t - t_*) = 0,2,4, \dots 100$.   } \label{fig:free-spectra}
      
      \begin{center} 
      \includegraphics[width=7.8cm]{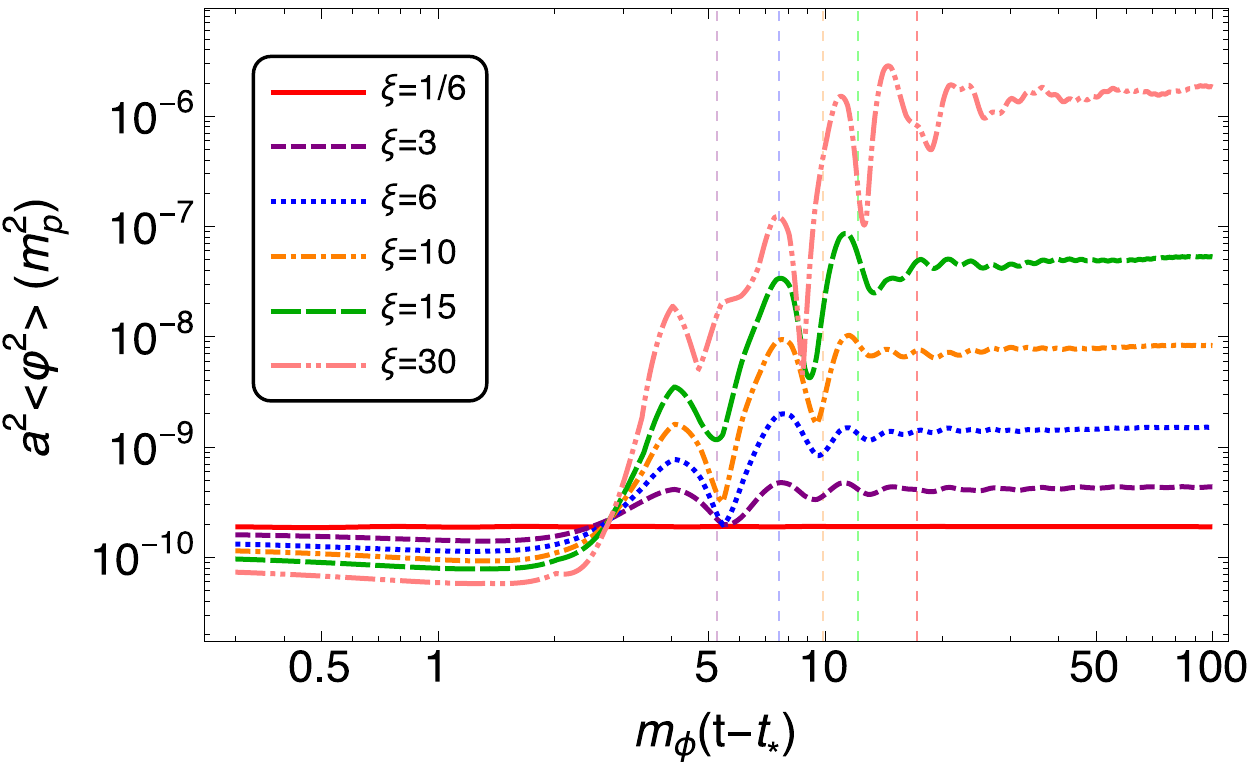}\hspace{0.6cm}
            \includegraphics[width=8.2cm]{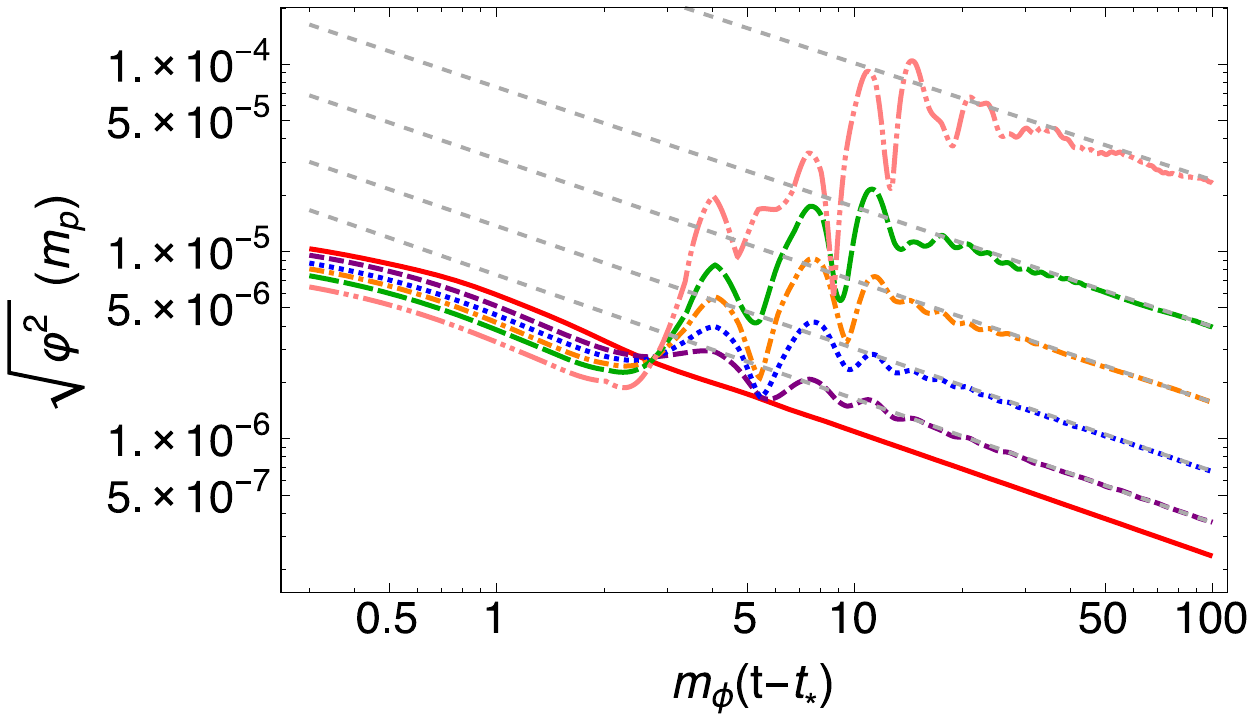}
      \end{center}
      \caption{Left: The Higgs conformal amplitude $a^2 \langle \varphi^2 \rangle$ obtained from lattice simulations, for the non-interacting case (Section~\ref{sec:iv}) with the Higgs-curvature couplings $\xi=1/6,3,6,10,15,30$. The dashed, vertical lines indicate the estimated time $t_{\rm res}$. Right: The root mean square of the Higgs physical amplitude $\langle \varphi^2 \rangle$ for the same couplings. We indicate in dashed lines the corresponding fit (\ref{eq:higgs-ampdecay}) for the late-time dynamics.}    \label{eq:rms-nolambda}    
      
\end{figure*}

In this Section, we study the case of a non-interacting scalar field, i.e.~we solve only the first equation in Eqs.~(\ref{eq:eom-fulleom1})-(\ref{eq:eom-fulleom}), setting $\lambda = g_s = 0$. Although this is obviously not a physical case, it will be helpful to understand our later results better when we include the Higgs self-interactive potential. Thus we consider now a 4-component Higgs field, coupled to the spacetime curvature through the term $\xi R \Phi^\dag\Phi$, with $R[\phi, \dot{\phi}]$ evolving due to the oscillating inflaton. We have done several lattice simulations of this system, varying the coupling $\xi$ within the range $\xi \in [4, 70]$. 

We show in Fig.~\ref{fig:free-spectra} the spectra of the Higgs field for the particular cases $\xi=5$ and $\xi=30$. In both panels, the red color corresponds to early times, while dark blue/purple corresponds to late times. In these spectra, a cutoff has been put in the distribution of initial fluctuations at the scale $k_c$, as indicated in Eq.~(\ref{eq:init-spectra}). The value of $k_c$ has been estimated from a previous set of lattice simulations without a cutoff, in which we see that, for $k > k_c$, the Higgs excitation due to the tachyonic resonance is negligible. Both spectra grow very fast, saturating eventually at a time $t \approx t_{\rm res}$, defined below. Naturally, the spectra grows several orders of magnitude more in the $\xi\approx 30$ case (right panel in Fig.~\ref{fig:free-spectra}) than in the $\xi \approx 5$ case (left panel in Fig.~\ref{fig:free-spectra}), as the tachyonic effect is stronger in the first case.

In Fig.~\ref{eq:rms-nolambda} we show the Higgs conformal and physical amplitudes as a function of time, averaged over the whole volume of the lattice, for the couplings $\xi=3,6,10,15,30$. We remind that this plot is for a four-component Higgs field, while for a single component we have $\langle \varphi_n^2 \rangle \approx \langle \varphi^2 \rangle /4 $ for each $n=1,2,3,4$. 

We expect the Higgs excitation to end when $q \lesssim 1$; see Eq.~(\ref{eq:q-term}). Taking $q = 0.2$ as the condition signaling the end of the tachyonic resonance regime, we find, using Eq.~(\ref{eq:PHI-term}), that the time $t_{\rm res}$ it takes to switch off the resonance, is
\be m_{\phi} (t_{\rm res}-t_*) \approx 1.58 \sqrt{4 \xi - 1} \ . \ee
 In the figure we indicate this time with vertical dashed lines. We see that for $t \lesssim t_{\rm res}$ particle creation is exponential, and the greater the $\xi$, the stronger the growth of the conformal Higgs amplitude $h = a \varphi$. However, as we approach $t \approx t_{\rm res}$, the Higgs excitation stops. From then on, the dynamics of the Higgs field is dominated by the expansion of the universe. More specifically, we have found that the late-time behaviour of the Higgs amplitude is
\be \langle | \varphi (t) |  \rangle \sim (m_{\phi} t)^{-(0.64 \pm 0.03)} \ , \hspace{0.3cm} m_{\phi} t \rightarrow  \infty \ , \label{eq:higgs-ampdecay}\ee
where the particular numerical value of the exponent depends on the value of $\xi$ considered. We indicate this in the left panel of Fig.~\ref{eq:rms-nolambda} with dashed lines. As expected, Eq.~(\ref{eq:higgs-ampdecay}) indicates that $\langle \varphi \rangle \propto a^{-1} (t)$. We have found that a rough estimate for the Higgs amplitude for late times is
\bea \langle |\varphi (t) | \rangle  &\approx & e^{s(\xi)} (m_{\phi} t)^{-p(\xi)} m_p \ ,  \nonumber \\*
s(\xi) &\equiv & -12.1 + 0.17 \xi + 0.00046 \xi^2 \ , \nonumber \\*
p (\xi) &\equiv & 0.67 - 0.0048 \log \xi - 0.0017 (\log \xi)^2 \ ,
\label{eq:fit-freecase}
\eea
where the first factor accounts for the initial excitation of the Higgs modes, and the second accounts for the later energy dilution.

Before we move on, it is important to note that, as we decrease $\xi$, the amplitude of the excited IR modes decreases significantly, being comparable to the amplitude of the (non-excited) UV modes for very low couplings. This signals that the lattice simulations cannot be trusted for these low couplings, because there is no significant excitation of the Higgs field over the initial vacuum fluctuations. Correspondingly, for these low couplings, the contribution of the UV modes to the Higgs amplitude becomes increasingly important, and hence its value can depend strongly on where we put the cutoff $k_c$ of the initial fluctuations. Therefore, there is a minimum value $\xi$ for which we can trust the lattice simulations. In this paper we have determined this condition as $\frac{\langle \varphi (t_{\rm res}) \rangle}{\langle \varphi_* \rangle} \frac{a (t_{\rm res})}{ a_* } > 2 $, which means basically that the contribution to the Higgs amplitude from the Higgs excitation, is greater than the one from the Higgs initial vacuum fluctuations. With this, we find that we cannot trust simulations with $\xi \lesssim 4$.

\section{Simulations with an unstable potential}
\label{sec:v}

Let us now move to simulations with the full Higgs potential~(\ref{eq:higgs-potential}), including
the four components of the Higgs field but yet without including gauge interactions. All the results of this Section have been obtained with lattice cubes of $N^3 = 256^3$ points, and minimum momentum $p_{\rm min} = 0.18 m_{\phi}$.

To get a qualitative understanding of the dynamics, let us recall the linearised equation of motion (\ref{eq:higgs-mode-eq}) for the Higgs field modes $h_k \equiv \varphi_k a^{3/2}$.
For high Higgs field values, $\varphi > \varphi_0$, the self-coupling is negative $\lambda (\varphi) < 0$, and therefore the interaction term tends to increase the Higgs field value, and  induce a transition to the negative-energy vacuum. The more the Higgs field has been amplified by the tachyonic resonance, the 
faster the instability is. On the other hand, because the Ricci scalar remains larger time positive than negative during each inflaton oscillaton, the non-minimal coupling term $\xi R(t)$ effectively creates a potential barrier than resists this increase. The amplitude of the curvature term decays as $\xi R(t) \propto a^{-3} (t) \propto t^{-2}$, so it becomes however gradually less important. If it contrarests the instability until the Higgs field amplitude has decreased below the barrier scale $\varphi<\varphi_0$, then the Higgs field remains stable throughout the entire evolution. Because the amplification by the tachyonic resonance depends exponentially on the non-minimal coupling $\xi$ [see Eq.~(\ref{eq:fit-freecase})], whereas the effective barrier due to $\xi$ depends on it only linearly, one expects that for high $\xi$, the instability takes place faster, and for low enough $\xi$ it is prevented completely.

\begin{figure}
      \begin{center} 
      \includegraphics[width=8.2cm]{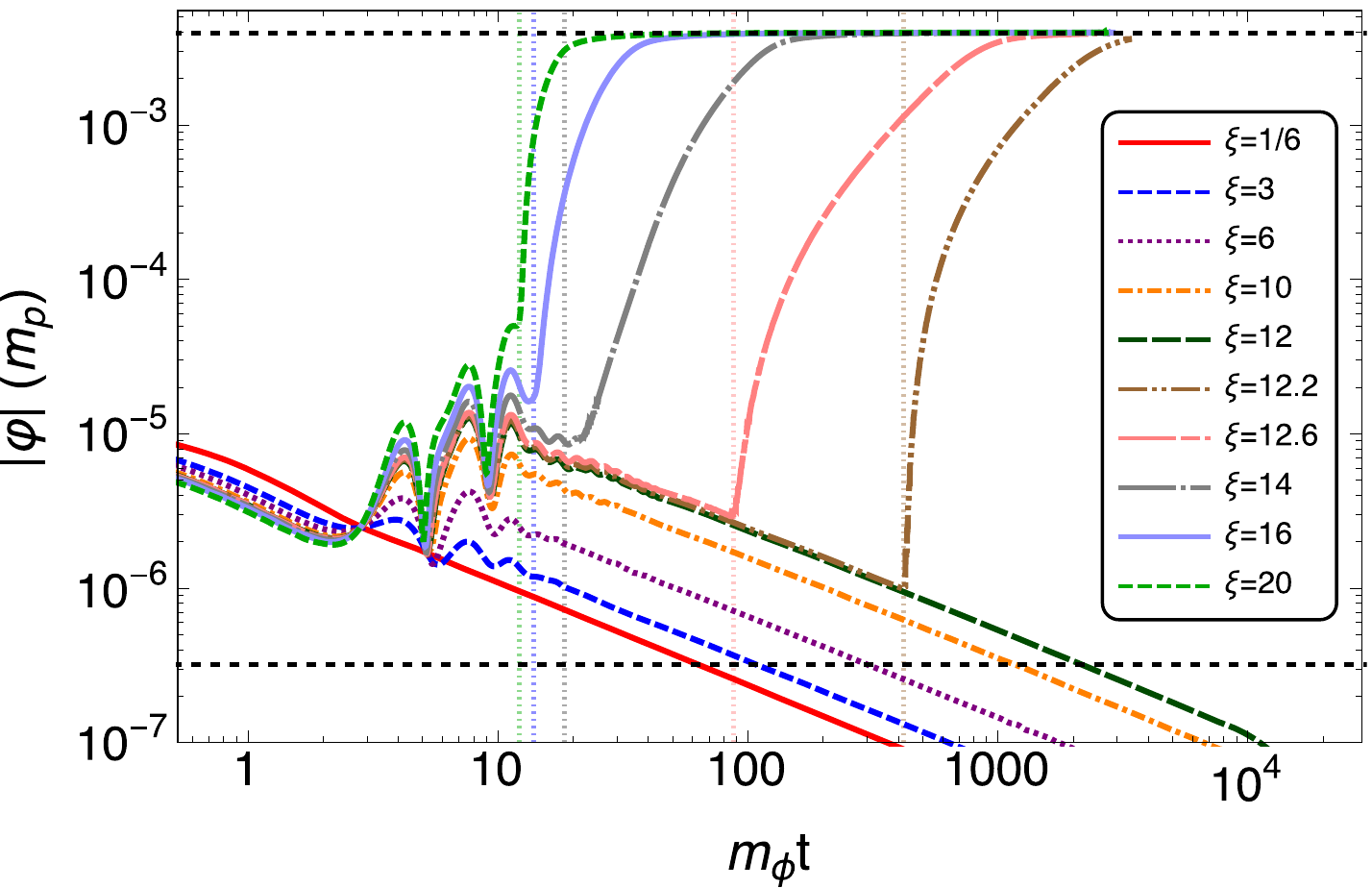}
      \end{center} 
      \caption{
The volume-average value of the Higgs field $|\varphi| = \sqrt{\sum_n \varphi_n^2 }$ obtained from lattice simulations
with unstable potential (Section~\ref{sec:v}), for the top quark mass $m_t=172.12 {\rm GeV}$. Each line represents simulations with a different value of $\xi$.  For the cases in which the Higgs field develops an instability, the vertical dashed lines indicate the instability time $m_{\phi} t_i$. The two dashed horizontal lines indicate the position of the barrier $\varphi_+$ estimated for this potential, and the (modified) high-amplitude, negative-energy vacuum $\varphi_v$.}
      \label{fig:inst-time}
\end{figure}

Fig.~\ref{fig:inst-time} shows the volume-averaged amplitude of the Higgs field $\langle |\varphi| \rangle$ as a function of time, for different choices of the Higgs-curvature coupling $\xi$, obtained directly from lattice simulations. In this Figure, we have used the running of the potential corresponding to the top quark mass $m_t = 172.12~{\rm GeV}$, see Fig.~\ref{fig:lambda-run}. This potential has the barrier at $\varphi_+ \approx 7.8 \times 10^{11} {\rm GeV}$. We can see that, for initial times $m_{\phi}(t-t_*) \lesssim 10$, the amplitude grows (in an oscillating way) due to the Higgs tachyonic resonance regime, as described in Section \ref{sec:ii}. 

In Fig.~\ref{fig:inst-time} we see that for high values of the non-minimal coupling, $\xi\ge 16$, the Higgs field becomes unstable during the tachyonic resonance, triggering a transition to the high-energy vacuum $\varphi = \varphi_v$. For lower values of the non-minimal coupling, the tachyonic resonance ends before the Higgs has become unstable. After this the behaviour is initially similar to the free field case discussed in Section~\ref{sec:iv}: the system settles in a quasi-stationary state in which the field amplitude gradually decreases due to the expansion of space. In the intermediate range of couplings, $12.2\le \xi\le 14$, the instability eventually takes place, at a time that we denote by $t_i$. We indicate this with a vertical dashed line in Fig.~\ref{fig:inst-time}. 

\begin{figure}
      \begin{center} 
			\includegraphics[width=8.2cm]{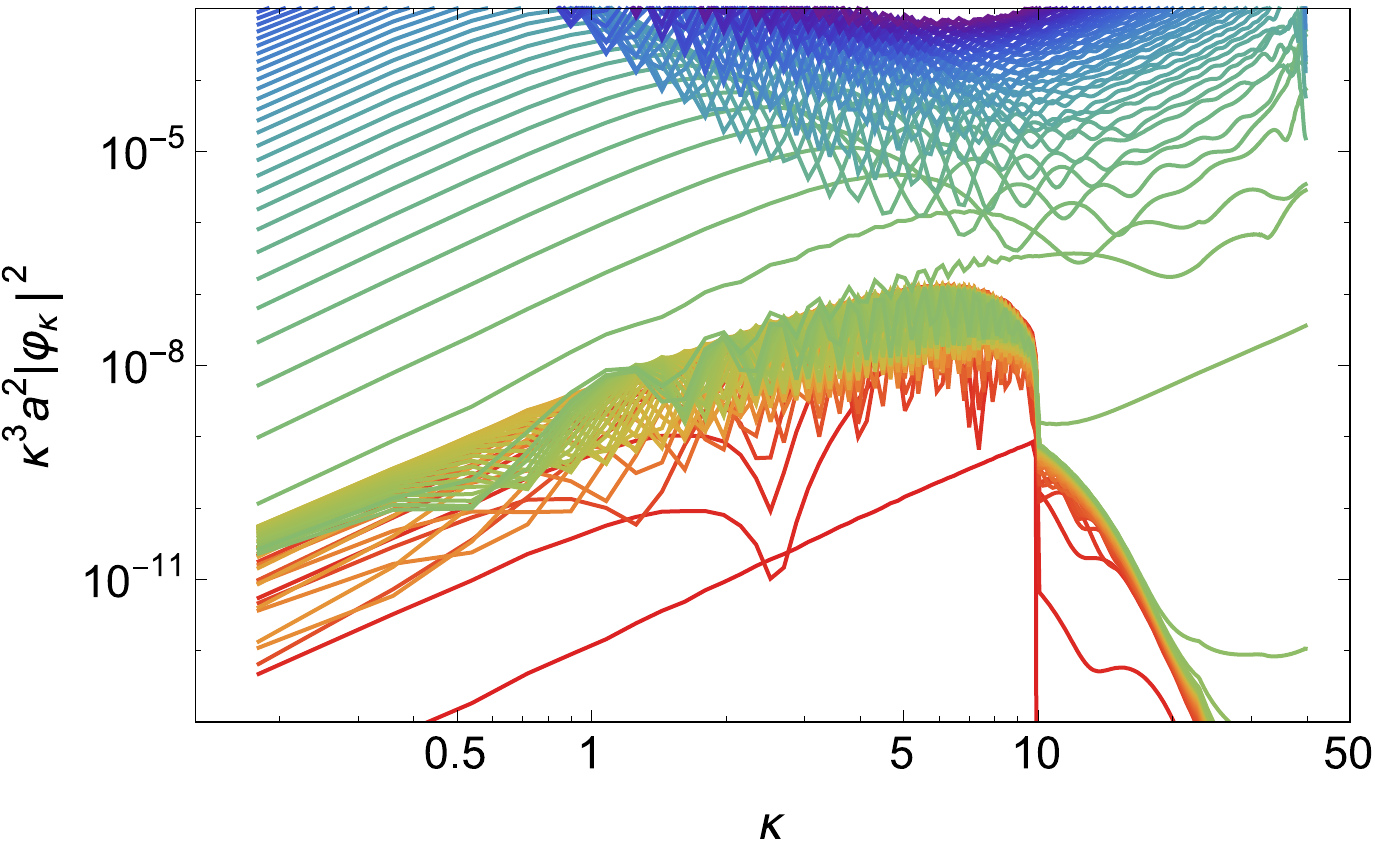}
      \end{center} 
      \caption{
Higgs field spectra $\kappa^3 a^2 |\varphi_k|^2$ as a function of $\kappa = k / m_{\phi}$ in the presence of a unstable potential (Section~\ref{sec:v}), for $\xi=12.2$ and $m_t=172.12\ {\rm GeV}$. The spectra is depicted at times $m_{\phi} (t - t_*) = 0,10,20...$, going from early times (red) to late times (dark blue).}
      \label{fig:inst-spectrum}
\end{figure}

\begin{figure}
      \begin{center} 
			\includegraphics[width=8.2cm]{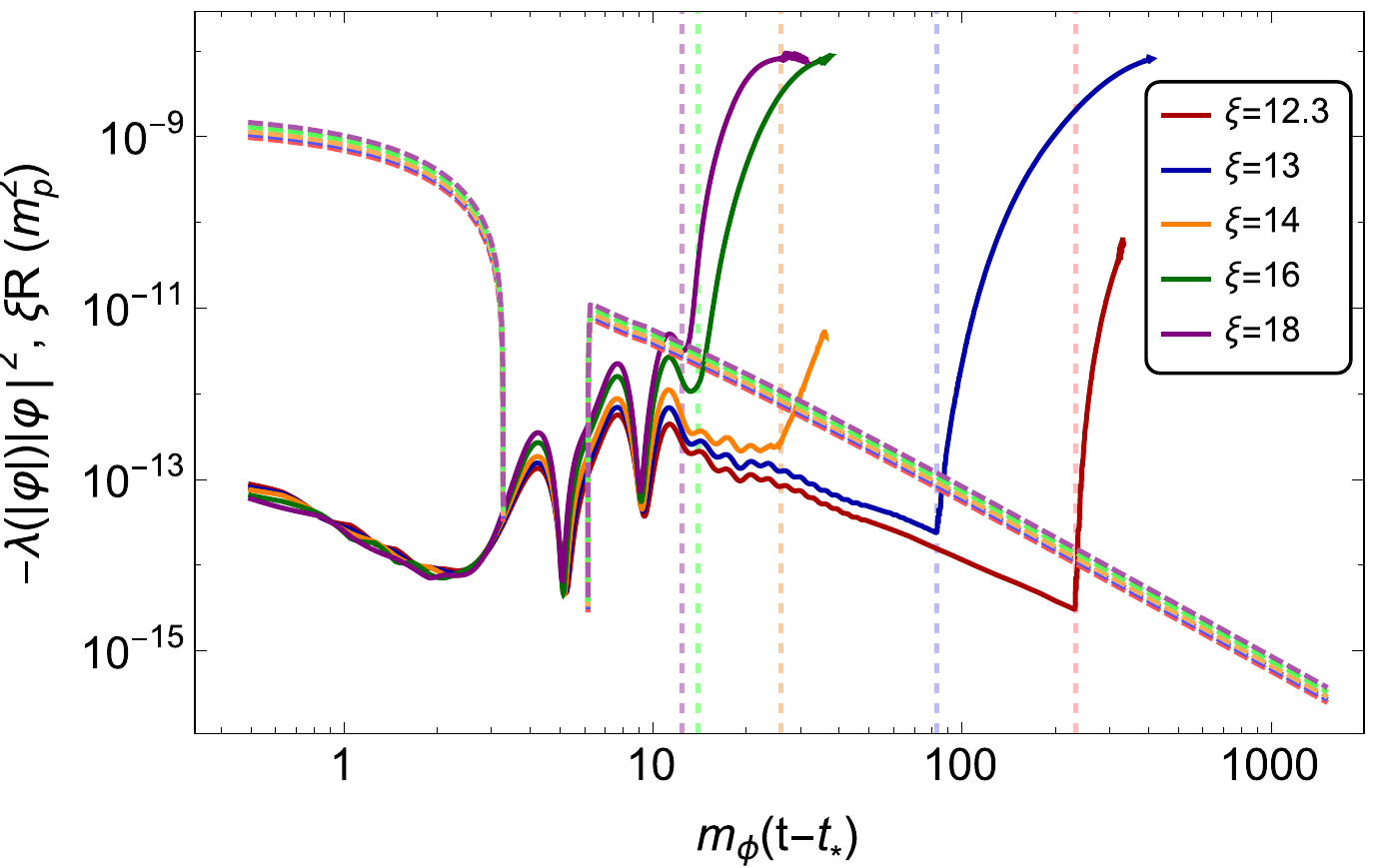}
      \end{center} 
      \caption{We show, for $m_t = 172.12 {\rm GeV}$ and different values of $\xi$ (Section \ref{sec:iv}), the time-evolution of the terms $\xi R$ (dashed lines) and $- \lambda (|\varphi|) \varphi^2$ (continuous lines). For $\xi R$, we plot an oscillation-average to compare both terms more easily. These are the terms that appear in the Eq.~(\ref{eq:higgs-mode-eq2}) for the field modes. We also plot, with vertical lines, the corresponding time $m_{\phi} t_i$ at which the Higgs becomes unstable.}
      \label{fig:term-comparison}
\end{figure}

For $\xi\le 12$, the field amplitude eventually decreases below the potential barrier, $\varphi<\varphi_+$. By this time, the barrier stabilises the field, and therefore the instability does not take place at all. This demonstrates that physically the instability is due to the tachyonic resonance. Even though the amplitude of the initial vacuum fluctuations is higher than the barrier scale, it is not high enough to lead to an instability before it is damped to safe values by the expansion of the universe. From the spectra shown in Fig.~\ref{fig:inst-spectrum} we can see that the infrared modes have to be amplified by roughly three orders of magnitude by the tachyonic resonance in order for the instability to take place. In particular, this means that the use of classical field theory simulations is well justified in this case.

We can explain the triggering of the Higgs instability in terms of the balance between the terms $\xi R$ and $-\lambda (\varphi) \langle \varphi ^2 \rangle$ that appear in the EOM of the field modes,  Eq.~(\ref{eq:higgs-mode-eq2}). We have plotted in Fig.~\ref{fig:term-comparison} the time evolution of these two terms for $m_t = 172.12 {\rm GeV}$ and different values of $\xi$. Although $\xi R$ is periodically oscillating between positive and negative values, the resulting oscillation average is always positive. We observe that initially, the first term dominates over the second, but as commented, when (if) the absolute value of the second term becomes of the same order of magnitude as the first one, the Higgs field becomes unstable. This can happen during the initial regime of tachyonic resonance, or later on due when the resonance is already switched off, as $R \propto 1/a^3$ whereas $\langle \varphi^2 \rangle \propto 1/a^2$.

In conclusion, as expected, we can define a critical coupling, $\xi_c \approx 12$ for $m_t = 172.12$ GeV, so that for $\xi \lesssim \xi_c$ the Higgs field is always stable, while for $\xi \gtrsim \xi_c$ the field becomes unstable at a certain time $m_{\phi} t_i$, whose numerical value decreases as $\xi$ gets greater. This general picture also applies for other values of the top mass. If we take the top quark mass a bit higher, $\varphi_+$ is lower, and hence the Higgs field takes a much longer time to settle on the safe side of the potential barrier. Because of this, the larger the mass $m_t$, the lower the value of the critical coupling $\xi_c$.

The order of magnitude fit of the time dependence of the amplitude obtained for the free case in Eq.~(\ref{eq:fit-freecase}) also holds quite well in the self-interacting scenario, before the instability takes place. This indicates that the effect on the Higgs dynamics of $\lambda$ is not very important before the transition to the high-energy vacuum takes place. Inverting this fit, we can find an order-of-magnitude estimate of the time $t_o$ at which we recover $\lambda (\varphi) > 0$, 
\be m_{\phi} t_{\rm o} \approx (\varphi_{\rm o} m_p^{-1} e^{-s (\xi)}) ^{-\frac{1}{p(\xi)}} \ , \label{eq:time-crossing}\ee
where $\varphi_{\rm o}$ is given in Table \ref{table:higgspot}. For $\xi \approx 5$, this gives $m_{\phi} t_{\rm o} \approx \mathcal{O} (10^{2,4,5,6,7})$ for top quark masses $m_t = 172.12, 172.73, 173.34, 173.95, 174.56~ {\rm GeV}$ respectively.

\begin{figure*}
      \begin{center} 
      \includegraphics[width=12cm]{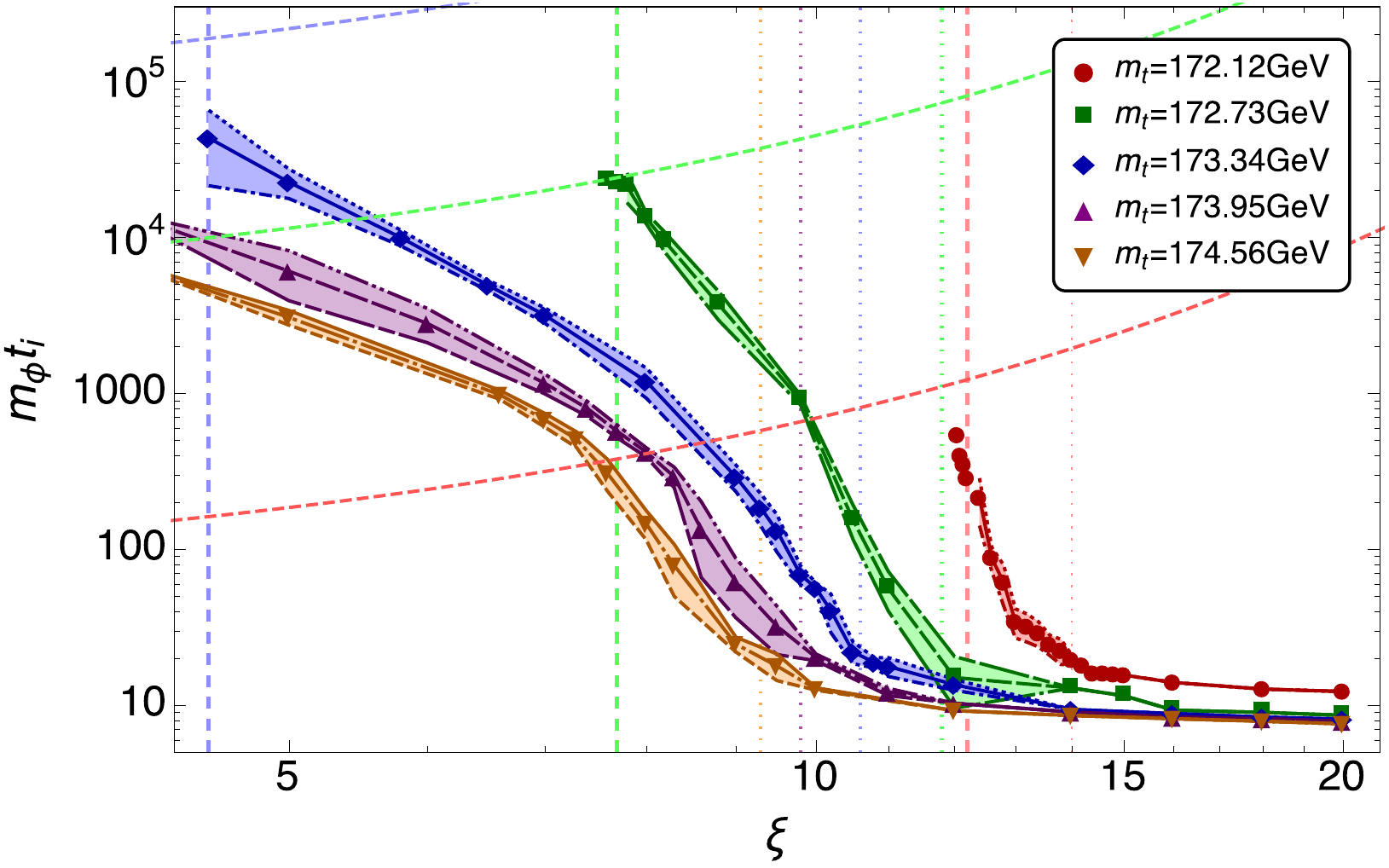}\hspace{0.5cm} \hspace{0.5cm}
      \end{center} 
      \caption{The instability time $m_{\phi} t_i$ at which the Higgs field develops an instability and decays to the true negative-energy vacuum, as a function of the Higgs-curvature coupling $\xi$ (Section \ref{sec:v}). These results are obtained directly from lattice simulations. Each of the five lines correspond to the five different interpolating potentials, corresponding to the top quark masses $m_t =172.12 {\rm GeV}$ (red), $m_t=172.73 {\rm GeV}$ (green), $m_t = 173.34 {\rm GeV}$ (blue), $m_t = 173.95 {\rm GeV}$ (purple), and $m_t = 174.56 {\rm GeV}$ (brown). The dashed vertical lines indicate the position of the critical couplings $\xi_{c}^{(1)}$, while the dotted lines indicate the position of the couplings $\xi_{c}^{(2)}$, see Table \ref{table:xicritical}. For each data point, we have done several lattice simulations corresponding to different realizations of the initial Higgs field conditions, see bulk text. The points indicate the average value $m_{\phi} t_i$, while the envelope of each of the lines indicates the standard deviation $\sigma \equiv N^{-1/2} \sqrt{ \sum_i (x_i - \bar{x}_i )^2 }$. For data points with $\xi \approx \xi_c$, only some of the ten simulations do not become unstable, and hence we do not show the deviation in these cases. For $\xi \lesssim \xi_{c}^{(1)}$ all simulations are always stable (i.e. $m_{\phi} t_i = \infty$), and hence data points are not drawn.}
\label{fig:plot-ti}
\end{figure*}

We show in Fig.~\ref{fig:plot-ti} the instability time as a function of $\xi$ obtained from our lattice simulations.  We have observed that the specific value of $m_{\phi} t_i$ depends on the particular random realization of the Higgs field initial conditions in Eq.~(\ref{eq:init-spectra}), so for each point, we have done several simulations for different realizations of the initial conditions (this is discussed in more detail in Section \ref{sec:vA}). Points indicate the average value of $m_{\phi} t_i$, while the shadow region surrounding each of the curves indicates the standard deviation.

The behavior of the five curves with $\xi$ is quite similar. In all curves, we can identify two critical values, $\xi_{c}^{(1)} \equiv \xi_c$ and $\xi_{c}^{(2)}$, which are identified in the Figure with dotted and dotted-dashed vertical lines, and indicated in Table \ref{table:xicritical}. The meaning of these values is as follows:

\begin{itemize}
\item $ \xi > \xi_{c}^{(2)}$: For these values, we observe that the Higgs field always develops an instability, at a time $m_{\phi} t_i \lesssim \mathcal{O} ( 10)$, quite independently on the value $\xi$ (at least for the cases we have simulated). This is seen as a plateau in the right part of the numerical curves shown in Fig.~\ref{fig:plot-ti}. Qualitatively, for this range of values, the Higgs field becomes unstable when it is still in the tachyonic resonance regime. One can see an example of this in Fig.~\ref{fig:inst-time} for $m_t = 172.12 ~{\rm GeV}$: For the cases $\xi=16,18,20$, which verify $\xi> \xi_{c}^{(2)} \approx 14$, the Higgs becomes unstable in the oscillatory regime, while for $\xi=12,14$, with $\xi \lesssim \xi_{c}^{(2)}$, the instability is developed when the resonance has already finished.

\item $\xi_{c}^{(1)} < \xi < \xi_{c}^{(2)}$: For these values, the Higgs field also develops an instability, but this happens only after the tachyonic resonance has ended. For these values, the instability time $m_{\phi} t_i$ depends very strongly on the value $\xi$. A change of few units in $\xi$ changes $m_{\phi} t_i$ in several orders of magnitude. 

\item $\xi < \xi_{c}^{(1)}$: Finally, for these values, we observe that the Higgs field is always stable, coming back to the safe side of the potential without having become unstable.

\end{itemize}

We indicate the values of both $\xi_{c}^{(1)}$ and $\xi_{c}^{(2)}$ for the cases $m_t=173.34, 172.73, 172.12 ~{\rm GeV}$ (blue, green, and red curves, respectively) in Table \ref{table:xicritical}. Note that, as expected, as we increase the value of the top-quark mass, the position of the barrier in the Higgs potential moves to smaller field values, and hence the initial distribution of the Higgs field is much deeper in the negative-energy region. Because of this, $\xi_{c}^{(1),(2)}$ are lower, and the Higgs field takes much longer to enter into the safe side of the potential.  Let us note that the identification of these critical values is not unambiguous, and in particular, for couplings near the critical one $\xi \approx \xi_c^{(1)}$, we observe that depending on the specific realization of the initial conditions, the Higgs may or may not become unstable. This source of uncertainty is indicated with a $\pm$ sign in Table~\ref{table:xicritical}. Finally, let us note that our technical definition of the second critical coupling is such that for $\xi > \xi_{c}^{(2)}$, we have $m_{\phi} t_i < 20$.

\begin{table}  
\begin{center}
  \begin{tabular}{ | c | c | c | c |  }
    \hline
    $m_t ({\rm GeV})$ & $ \xi_{c}^{(1)}$ & $ \xi_{c}^{(2)} $  \\ \hline
    $172.12$  & $12.2 \pm 0.2$ & $ 14$ \\  \hline
    $172.73$  & $7.7 \pm  0.1$ & $ 11.8$  \\  \hline
    $173.34$  & $4.3 \pm  0.2$ & $ 10.6$  \\  \hline
    $173.95$  & $<4.0$ & $ 9.8$  \\  \hline
     $174.56$  & $<4.0$ & $ 9.3$  \\  \hline
  \end{tabular}
\end{center} 
\caption{Higgs-curvature critical couplings $\xi_{c}^{(1)}$ and $\xi_{c}^{(2)}$, obtained from lattice simulations, for different values of the top quark mass. The error in $\xi_c$ signals the uncertainty with respect initial conditions. The meaning of this parameters is explained in the bulk text.}\label{table:xicritical}
\end{table}

The curved, dashed lines in Figure~\ref{fig:plot-ti} indicate the approximated time at which the Higgs enters into the safe side of the potential, using Eq.~(\ref{eq:time-crossing}). The idea is that at the critical coupling $\xi=\xi_c^{(1)}$, the curve for $m_{\phi}t_i$ obtained from the numerical simulations (bands in colors in Fig.~\ref{fig:plot-ti}) will meet approximately the corresponding dashed ones. We can see in Fig.~\ref{fig:plot-ti} that this works relatively well, taking into account that Eq.~(\ref{eq:time-crossing}) is only a rough estimation. 

In Fig.~\ref{fig:plot-ti} it can also be seen that for $m_t= 173.95~ {\rm GeV}$ and $m_t = 174.56~{\rm GeV}$, the instability curves do not meet their corresponding curved-dashed lines for $\xi \gtrsim 4$, which are the cases that we cannot study in the lattice as discussed at the end of Section \ref{sec:iv}. Hence, for these masses we can provide only the upper bound $\xi_c \lesssim 4$.

As a final comment, let us note that in all our simulations we made the inflaton to oscillate indefinitely, even though this is clearly not realistic.
The inflaton is expected to be coupled to other species which will eventually induce its decay due to parametric resonance effects
at a time that we denote by $t_{\rm br}$, where the label $_{\rm br}$ stands for the backreaction from the decay products of the inflaton. After this time, the energy density is no longer dominated by a coherently oscillating scalar field, and therefore Eq.~(\ref{eq:ricci-eom}) is no longer valid. This puts an end to the tachyonic resonance regime of the Higgs field. Therefore, the estimates for $\xi_c$ provided here will not be valid if $t_{\rm br} \lesssim t_{\rm res}$. For example, as seen in \cite{FitParamResonance}, if the inflaton is coupled to a single scalar field $\chi$ with coupling $g^2 \phi^2 \chi^2$, one finds $m_{\phi} t_{\rm br} \gtrsim 40$ for $g^2 \lesssim 6.9 \cdot 10^{-3}$, so that $t_{\rm br} \gtrsim t_{\rm res}$ for the values of $\xi$ considered here. Therefore, in this case, our bounds can be applied. 

\subsection{Dependence of lattice simulations on the Higgs number of components and initial conditions} \label{sec:vA}

We address now how our results depend on the position of the momenta cutoff in the spectra of initial conditions, as well as on the number of Higgs components we put in our simulations. 

\subsubsection{Dependence on Higgs initial conditions} 

We have explained previously how the initial conditions of the Higgs field are set throughout the lattice. Basically, we impose at initial time $t=t_*$ vanishing homogeneous modes $\varphi_n = 0$ ($n=1,2,3,4$), and then we add quantum fluctuations to each of the components. These fluctuations are imposed only up to a certain cutoff momentum $k_c$, so that for $k > k_c$ the fluctuations are set to zero. Also, the random nature of the initial conditions is implemented in the code through a pseudo-random number generator, so that different seeds produce different realizations for the initial conditions. 

\begin{figure}
      \begin{center} 
      \includegraphics[width=8.2cm]{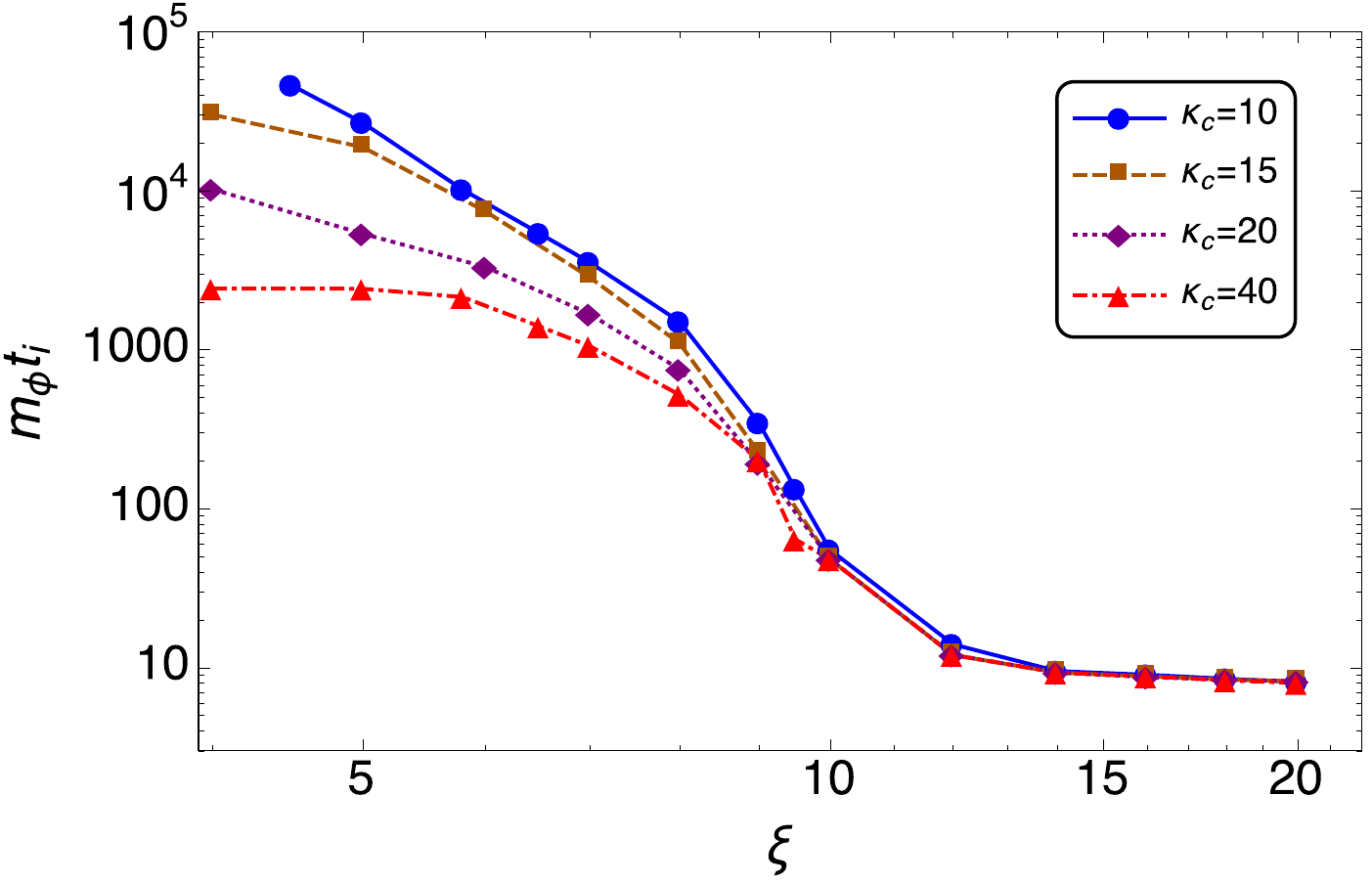}
                  \includegraphics[width=8.2cm]{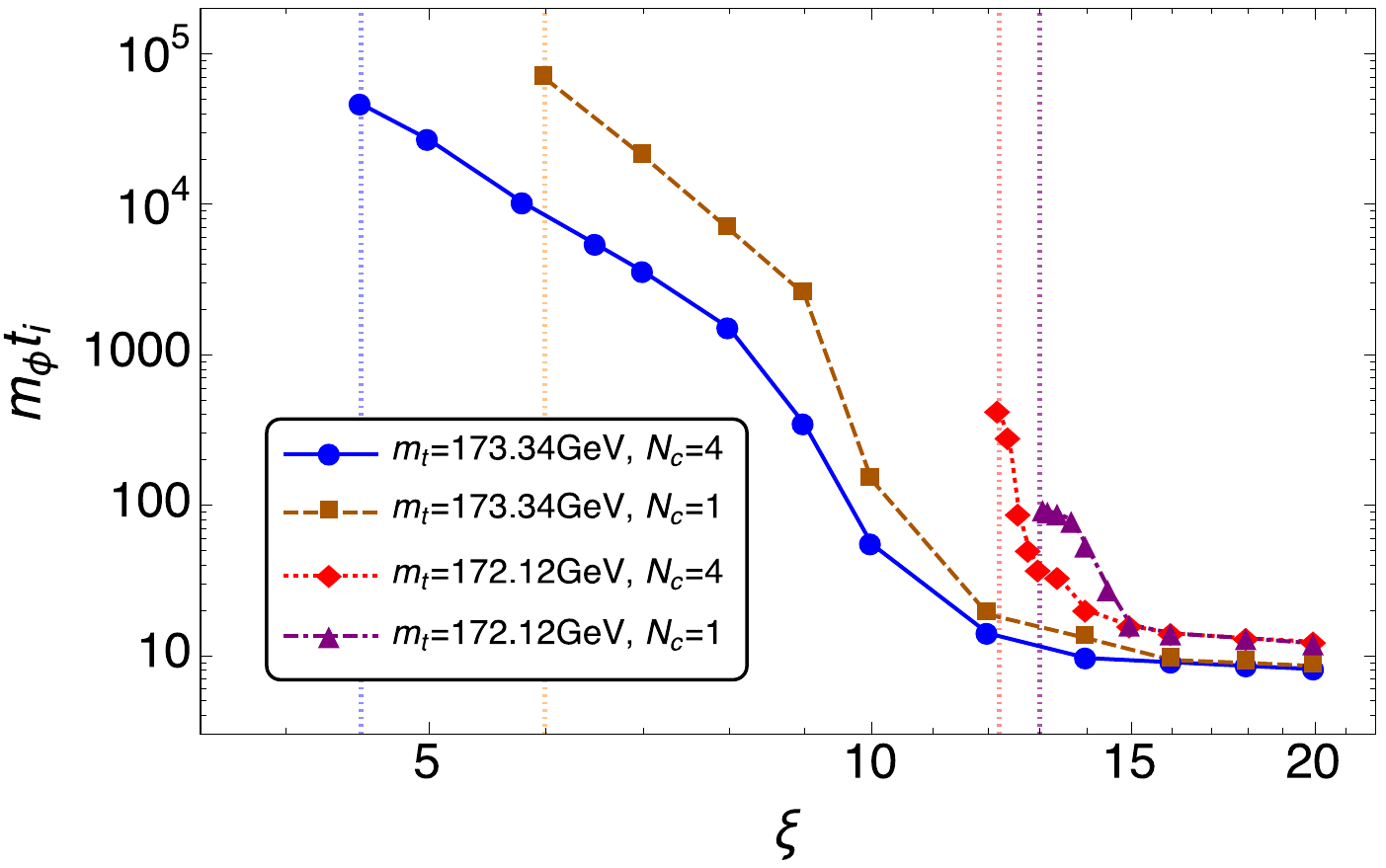}
      \end{center} 
      \caption{Top: The instability time $m_{\phi} t_i$ for the top quark mass $m_t = 173.34~{\rm GeV}$. Each curve corresponds to a different value of the cutoff of the initial fluctuations $\kappa_c \equiv k_c / m_{\phi}$, and each point corresponds to a particular lattice simulation. We depict here the interval $4 < \xi < 20$. Bottom:  The instability curves for $m_t=173.34~{\rm GeV}$ and $m_t=172.12~{\rm GeV}$  when we introduce a 4-component or a 1-component Higgs field. Each point corresponds to a single lattice simulation.}
      \label{fig:kcutoff-dependence}
\end{figure}

It is essential to fix the initial cutoff appropriately, so that the non-excited UV quantum modes, which cannot be treated in the lattice, are not excited as classical modes. In the results presented in Fig.~\ref{fig:plot-ti}, we have done several simulations with different initializations for each point. More specifically, for values $\xi < 5$, we have done ten simulations, five of them with $\kappa_c = 10$  ($\kappa_c \equiv k_c /m_{\phi} $), and the other five with $\kappa_c = 12$. We have also varied the seed in each of the ten simulations. This matches quite well the analytical estimation for the classical estimation of modes during tachyonic resonance given in Eq.~(\ref{eq:momenta-cutoff}). For values $5 < \xi < \xi_{c}^{(2)}$, the second set of five simulations has been done instead with cutoff $\kappa_c = 15$. Finally, for points $\xi > \xi_{c}^{(2)}$, we have done only four simulations (two with $\kappa_c = 10$ and two with $\kappa_c = 15$), because for these points the dependence of our results on the initial conditions is negligible. 

The top panel of Fig.~\ref{fig:kcutoff-dependence} shows how the instability curves change for different choices of the initial cutoff $\kappa_c$ for the particular case $m_t = 173.34 ~{\rm GeV}$. The inclusion of UV modes in the lattice beyond the physical cutoff, makes larger the Higgs amplitude $\langle h^2 \rangle$, so that the negative $\lambda \langle h^2 \rangle /a^3 $ term in Eq.~(\ref{eq:higgs-mode-eq2}) is enhanced, and hence reduces the instability time $m_{\phi} t_i$. For $\xi \gtrsim 10$, this effect is negligible, because as we saw in Section \ref{sec:iv}, the amplitude of the excited IR modes dominates over the UV ones, but it becomes increasingly important as $\xi$ diminishes. As we decrease the coupling, the UV modes become more relevant, and if they are not appropriately eliminated, their contribution can make the instability time wrongly smaller. At very low couplings, this is related to the invalidity of the lattice approach, as explained in the last paragraph of Section \ref{sec:iv}. 

Finally, let us note that, although the vacuum always becomes unstable for values $\xi > \xi_c^{(1)}$, the opposite condition $\xi<\xi_c^{(1)}$ does not guarantee stability. To show that, we would need to do $\sim e^{180}$ runs to account for the number of different causally disconnected patches of the Universe, and check that none of them leads to vacuum decay. This is not feasible, so we simply exclude parameters where vacuum decay happens in a typical run. Our sampling of initial conditions show in any case, that the statistical variance of the critical couplings, is much smaller than the critical couplings themselves. Therefore it is safe to expect that there would be full stability at some critical value slightly smaller than but of the order of the ones found. Besides, the uncertainty in $m_t$ propagates as a much larger effect than the statistical uncertainty in $\xi_c$.

\subsubsection{Dependence on Higgs number of components}

We now compare our results, in which we have taken the Higgs as a 4-component field ($N_c = 4$), with a similar set of lattice simulations with a 1-component field ($N_c=1$). 

We expect differences between the two scenarios for several reasons. The first one is that, if we include a 4-component field, the tachyonic mass is exciting 4 scalar fields instead of one. If we neglect at first the Higgs self-interaction term, this means $\langle \varphi_n^2 \rangle \approx \langle \varphi^2 \rangle /4 $ ($n=1,2,3,4$). 
Because of this, if we consider only a 1-component Higgs, the magnitude of the negative self-interaction term is being underestimated, and increases artificially the instability time $m_{\phi} t_i$ for a given coupling $\xi$, as well as the critical value $\xi_c$.

To check this, we show in the bottom panel of Fig.~\ref{fig:kcutoff-dependence} the dependence of the instability curve on the number of components, for the top quark masses $m_t=172.12~{\rm GeV}$ and $m_t = 173.34~{\rm GeV}$. We compare the cases $N_c=4$ (i.e. the case we have presented above), and $N_c = 1$. As expected, for the 1-component case the critical coupling $\xi_c$ increases slightly. For the $m_t = 173.34~{\rm GeV}$ case, we have $\xi_c \approx 6$ instead of $\xi_c \approx 4$, while for the $m_t = 172.12~{\rm GeV}$ we have $\xi_c \approx 13$ instead of $\xi_c \approx 12$. Apart from that, we see that the particular shape of the instability curve is significantly changed, meaning that the effect of the interaction between the different Higgs components is  relevant for the dynamics of the system.

\section{Simulations with gauge fields}
\label{sec:vi}

Until now, we have ignored the coupling of the Higgs field to the gauge bosons of the Standard Model. We now evaluate if the effects of this interaction modify significantly the results presented in the last Section.

Let us consider the following action in the continuum, which imitates the interactions of the Higgs with the four $SU(2)\times U(1)$ weak and hypercharge gauge bosons ($W_{1,2,3}$ and $Y$), in the Abelian approximation:
\bea S&=& - \int a^3 (t) d^4 x  \left( \frac{1}{4 g^2} \sum_{a=1}^3 W_{\mu \nu}^a W^{\mu \nu}_a + \frac{1}{4 g'^2} Y_{\mu \nu} Y^{\mu \nu} \right. \nonumber \\
&&  + (D_{\mu} \Phi)^{\dagger} (D^{\mu} \Phi ) + \lambda (\Phi^{\dagger} \Phi)^2 \Big) \ . \label{eq:action3gauge}
\eea
Here, $W_{\mu \nu}^a \equiv \partial_{\mu} W_{\nu}^a - \partial_{\nu} W_{\mu}^a$, $Y_{\mu \nu} \equiv \partial_{\mu} Y_{\nu} - \partial_{\nu} Y_{\mu}$, and $(D_{\mu})_{ij} \equiv \delta_{ij} (\partial_{\mu} - i Y_{\mu} /2 - i W_{\mu}^a / 2 )$. We take the ${\rm SU(2)}$ and ${\rm U(1)}$ couplings as constants with $g^2 \approx 0.3$, and $g'^2 \approx 0.3$ (corresponding to their value at very high energies, according to the SM renormalization group). This action describes correctly the Higgs-gauge fields interactions of the SM, as long as the non-linear interactions of the gauge fields among themselves (due to the truly non-Abelian nature of the SM symmetries) can be ignored. This is typically a good approximation as long as the gauge fields are not largely excited. 

As shown in \cite{Figueroa2015}, a system with $N > 1$ Abelian gauge-bosons can be equivalently described by a single effective gauge boson $A_{\mu}$. We can define the gauge boson field amplitudes in terms of the effective gauge boson as follows
\be Y_{\mu} \equiv \frac{g'^2}{g_s^2} A_{\mu} \ , \hspace{0.3cm} W_{\mu}^a \equiv \frac{g^2}{g_s^2} A_{\mu} \ ; \hspace{0.3cm} g_s^2 \equiv g'^2 + 3 g^2\,. \ ; \label{eq:super-boson}\ee
Substituting Eq.~(\ref{eq:super-boson}) in action (\ref{eq:action3gauge}), we recover action (\ref{eq:action-SM}) for an effective gauge boson with gauge coupling $g_s^2 \approx g'^2 + 3 g^2$. The effective gauge boson is then simply the sum of the original ones, i.e.~$A_{\mu} = Y_{\mu} + \sum_{a=1}^3 W_{\mu}^a$.

Naturally, what our lattice simulations do is to solve a discrete version of Eqs.~(\ref{eq:eom-fulleom}), which we provide in Eq.~(\ref{eq:eom-gaugediscrete}) of the Appendix. Details of how we derive this equations and the assumptions we made are provided in more detail there. The results we present in this section are based on lattice simulations with $N^3 = 128^3$ points, with a minimum infrared momenta $k_{\rm min} = 0.5 m_{\phi}$. This captures quite well the relevant range of momenta excited during the tachyonic resonance regime, for both the Higgs and the gauge fields. 

\begin{figure}
      \begin{center} 
      \includegraphics[width=8.2cm]{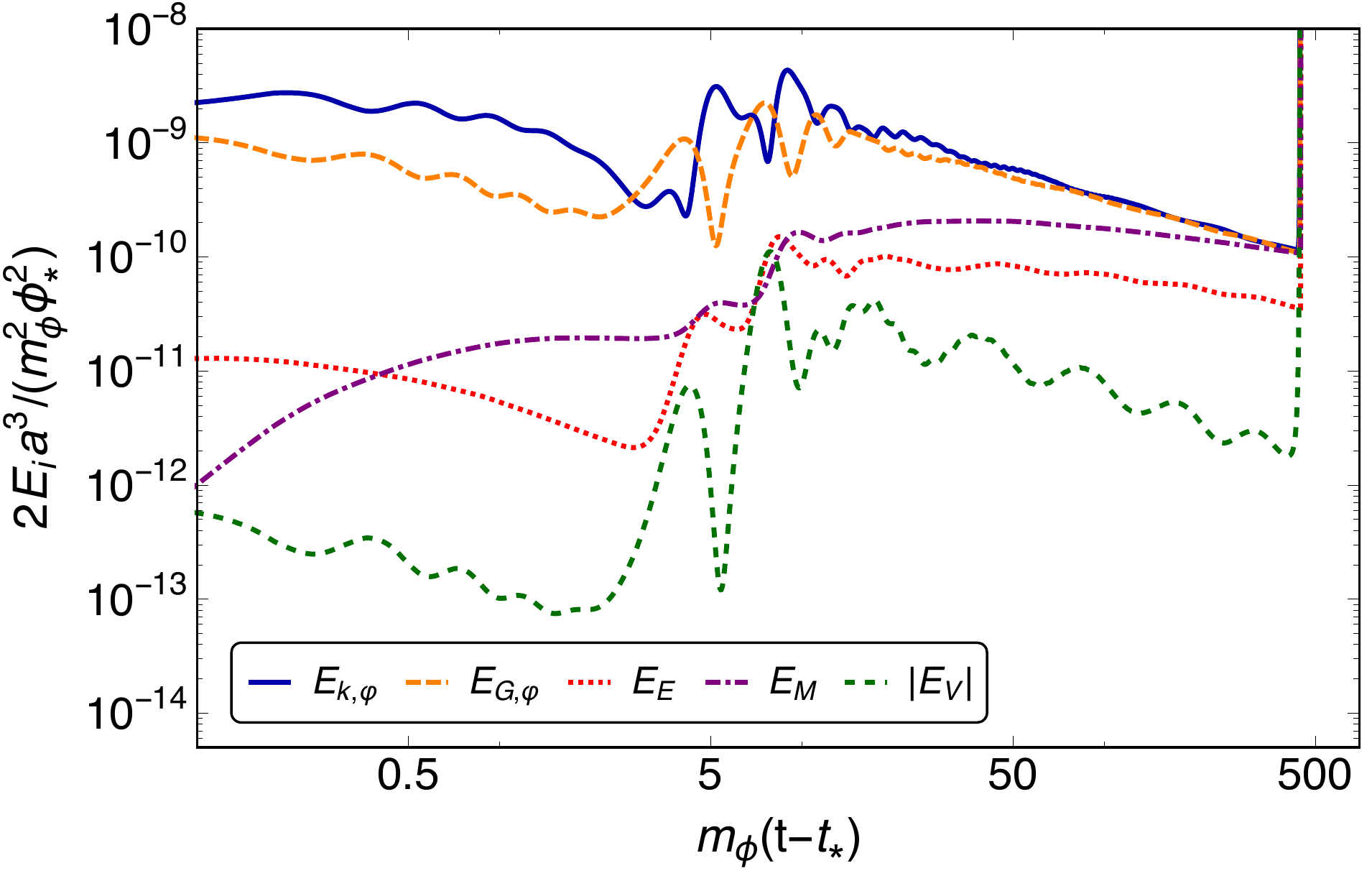}
      \end{center} 
      \caption{We plot, for $\xi=8$ and $m_t = 173.34 {\rm GeV}$, the different contributions to the energy density (\ref{eq:energy}) as a function of time (Section \ref{sec:vi}). The Higgs field becomes unstable around $m_{\phi} t_i \approx 500$.}
      \label{fig:energies}
      \begin{center} 
      \includegraphics[width=8.2cm]{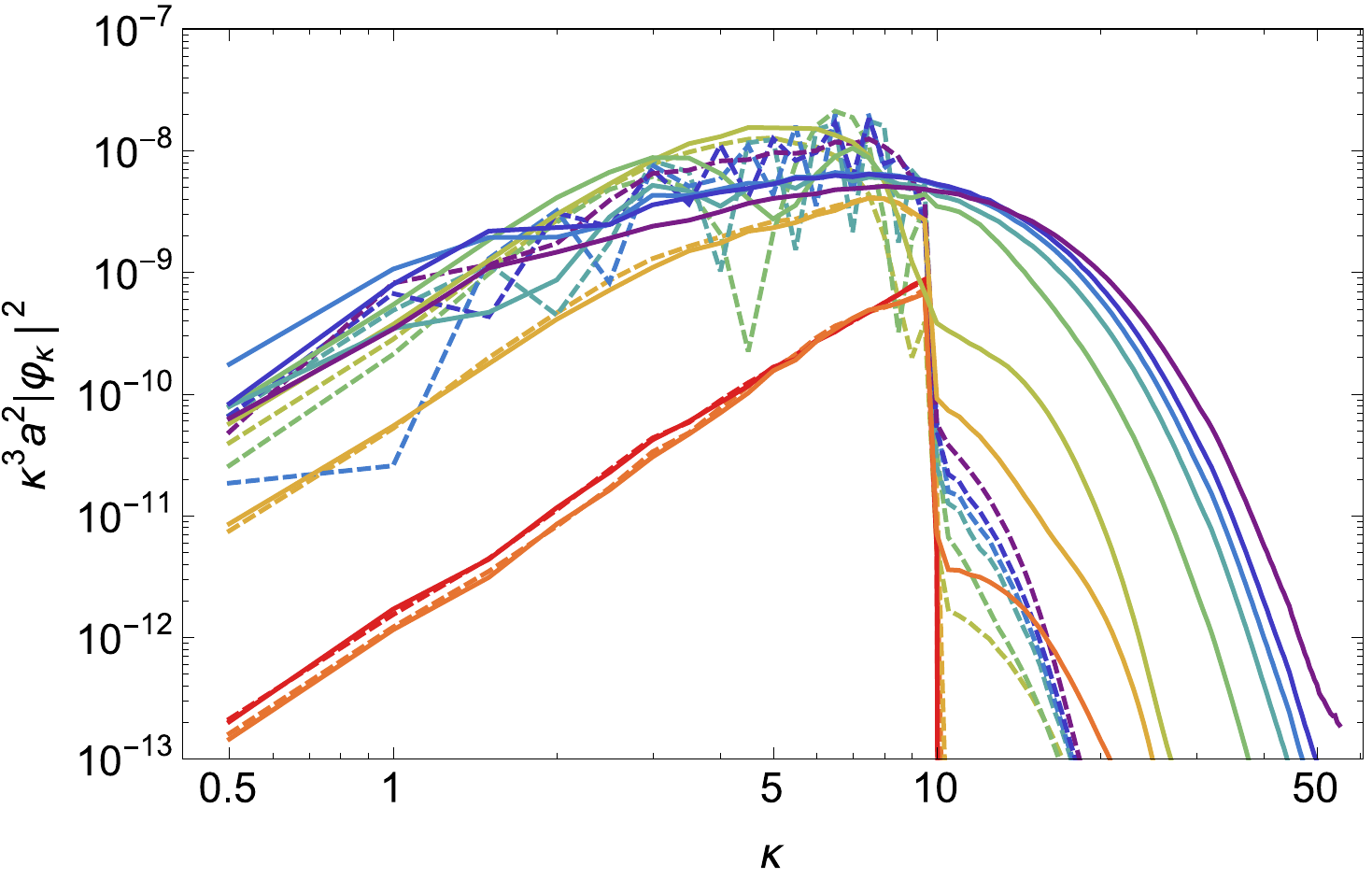}
      \end{center} 
      \caption{We plot the spectra of the conformal Higgs field for different times (Section \ref{sec:vi}). Continuous lines correspond to a Higgs coupled to gauge bosons ($g_s^2 =1.2$), while the dashed lines indicate the equivalent when such coupling is set to zero ($g_s^2 =0$). Here, we have chosen $m_t = 173.34 {\rm GeV}$ and $\xi = 8$. From early (red) to late times (purple), we have $m_{\phi} ( t - t_*) = 0,2,4,8,18,59,100,161,403$.}
      \label{fig:comparison-spectra}
\end{figure}

Let us try to quantify the energy transferred from the Higgs into the electroweak gauge bosons. Action (\ref{eq:action-SM}) can be written as $S = S_m + S_R$, with $S_R \equiv \int d^4 x \sqrt{- g} \xi R |\Phi |^2$ containing the Ricci-Higgs interaction term, and $S_m$ containing the other terms. We define the matter stress-energy tensor as $T_{\mu \nu}^{(m)} = \frac{2}{a^3} \frac{\delta S_m}{\delta g^{\mu \nu}}$. The energy density can then be written as
\bea T_{00}^{(m)} &=& \frac{1}{2} |\dot{\varphi} |^2 + \frac{1}{2 a^2} \sum_i |D_i \varphi |^2 + \frac{1}{2 a^2 } \sum_i F_{0i}^2  \nonumber \\*
&& + \frac{1}{2 a^4} \sum_{i,j<i} F_{ij}^2 + V(| \varphi |) \nonumber \\*
&& \equiv E_K^{\varphi} + E_G^{\varphi} + E_E + E_M + E_V \ . \label{eq:energy} \eea

\begin{figure*}
      \begin{center} 
      \includegraphics[width=11cm]{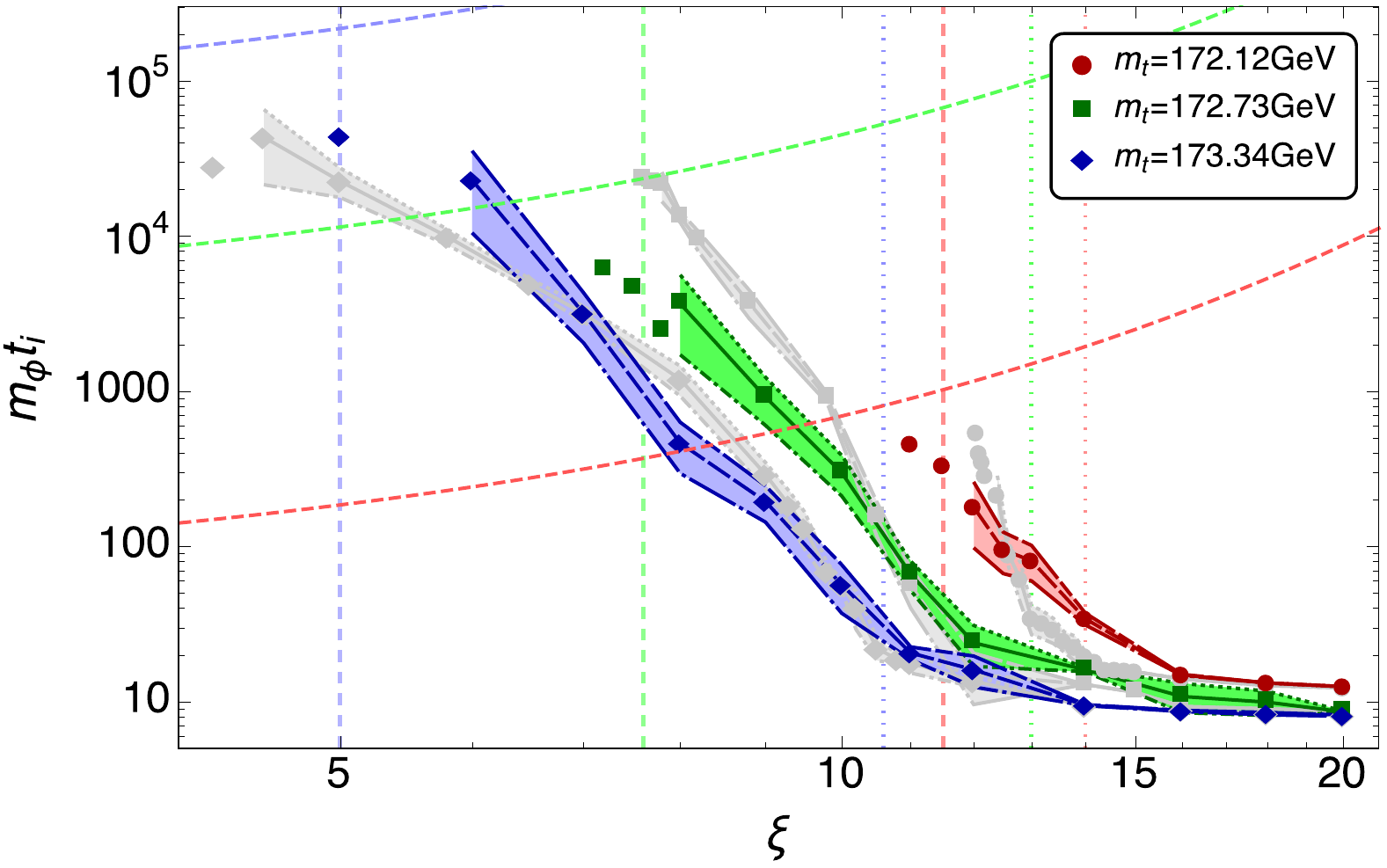}\hspace{0.5cm} \hspace{0.5cm}
      \end{center}
      \caption{ The instability time $m_{\phi} t_i$ as a function of $\xi$, obtained from the lattice simulations with both Higgs and gauge bosons (Section \ref{sec:vi}). We have depicted the cases for the top-quark mass $m_t=172.12 {\rm GeV}$ (red), $m_t = 172.73 {\rm GeV}$ (green), and $m_t=173.34 {\rm GeV}$ (blue). The three gray curves show the results, for these same three masses, of the lattice simulations with no gauge bosons incorporated (i.e. the curves of Fig.~\ref{fig:plot-ti}). As before, the dashed, and dotted-dashed vertical lines indicate the estimations $\xi_c^{(1)}$ and $\xi_c^{(2)}$ respectively, whose meaning is described in the bulk text of Section \ref{sec:v}, and the curved dashed lines show the estimation of Eq.~(\ref{eq:time-crossing}) for the three different top-quark masses.}
\label{fig:plot-tigauge}
\end{figure*}

\begin{table}  
\begin{center}
  \begin{tabular}{ | c | c | c | c |  }
    \hline
    $m_t ({\rm GeV})$ & $ \xi_{c}^{(1)}$ & $ \xi_{c}^{(2)} $  \\ \hline
    $172.12$  & $11.3 \pm 0.4$ & $ 15 $ \\  \hline
    $172.73$  & $7.4 \pm  0.3$ & $ 13$  \\  \hline
    $173.34$  & $5 \pm  0.5$ & $ 11$  \\  \hline
  \end{tabular}
\end{center} 
\caption{Higgs-curvature critical couplings $\xi_{c}^{(1)}$ and $\xi_{c}^{(2)}$, obtained from lattice simulations for different values of the top quark mass, in the presence of a coupling of the Higgs field to the gauge bosons. The meaning of this parameters is explained in the bulk text.}\label{table:xicritical-gauge}
\end{table}

We show in Fig.~\ref{fig:energies} the evolution of the different contributions to the energy density (\ref{eq:energy}) as a function of time, for the case $\xi = 8$. These energies have been divided by the inflaton energy $\sim \frac{m_{\phi}^2 \phi_*^2}{2 a^3 } $. We see that the Higgs and gauge fields energy is several orders of magnitude lower than the inflaton energy, which justifies neglecting their contribution to the Friedmann equation, as commented in Section \ref{sec:iii}. At late times, the Higgs kinetic and gradient energies evolve as $E_K^{\varphi}, E_G^{\varphi} \sim a^{-4}$, and thus eventually become subdominant with respect the magnetic energy.

We show in Fig.~\ref{fig:comparison-spectra} the time-evolution of the Higgs spectra in the presence of a gauge interaction with $g_s^2 = 1.2$, and compare it when such an interaction is not present ($g_s^2 = 0$). We clearly see that the gauge bosons have a very important backreaction effect on the Higgs field, propagating its spectra to the UV.

Finally, Fig.~\ref{fig:plot-tigauge} shows the instability time $m_{\phi} t_i$ as a function of $\xi$ obtained from lattice simulations, when we do include the coupling of the Higgs with the gauge bosons. We have simulated the cases $m_t = 172.12, 172.73,173,34 {\rm GeV}$, and compared with the results obtained in Section \ref{sec:v}, when we ignored such coupling. Although the instability curves are slightly different with respect to the case without gauge bosons, the values for the critical coupling $\xi_c^{(1)}$ and $\xi_{c}^{(2)}$ do not change significantly. We show these values in Table \ref{table:xicritical-gauge}.

In conclusion, our simulations demonstrate that the addition of gauge fields does not impact significantly in the post-inflationary dynamics of the system. The interaction of the Higgs with the electroweak gauge fields changes only marginally the results on the critical couplings $\xi_c^{(1), (2)}$. Besides, as we used an Abelian set-up, this also indicates that the addition of the truly non-Abelian gauge bosons will not change the above conclusion, as the non-linear nature of the non-Abelian gauge field interactions cannot stimulate further the gauge bosons. Quite on the contrary, the non-linear structure of non-Abelian interactions typically prevents the stimulation of the gauge fields up to the level of excitation that (linear) Abelian interactions allow for. 

\section{Summary and discussion}
\label{sec:vii}

In this work, we have studied the post-inflationary dynamics of the Standard Model Higgs with lattice simulations, in the case where it possesses a non-minimal coupling $\xi$ to gravity. This term is necessary for the renormalization of the theory in curved spacetime. We have assumed a chaotic inflation model with $m_{\phi}^2 \phi^2$ potential. We include the running of $\lambda (\varphi)$ in our simulations as a function of the value of the Higgs field at the lattice point. We have considered different runnings, corresponding to different experimental values of the top-quark mass. The running is such that it generates two vacua to the Higgs potential: one at $ \varphi \approx 0 $, and one at high-energies. With our lattice simulations, we have been able to obtain the critical coupling $\xi_c$ such that for $\xi \gtrsim \xi_c$ the Higgs field becomes unstable and decays into the negative-energy Planck-scale vacuum. Our lattice simulations also take into account the 4 components of the Higgs field and the cutoff of the spectra of initial fluctuations, which are necessary to correctly quantify the value of $\xi_c$. We have done two sets of lattice simulations; one with only the Higgs field, including the effective expansion caused by the post-inflationary dynamics of the inflaton; and another in which we also include the coupling of the Higgs to gauge bosons (modeled with an Abelian-Higgs-like approach). We have observed that the effect of the gauge bosons is not relevant for the Higgs post-inflationary dynamics.

The upper bounds in Tables~\ref{table:xicritical} and \ref{table:xicritical-gauge}, together with the estimation $\xi \gtrsim 0.06$ from the stability of the Higgs field during inflation~\cite{Herranen2014}, provide tight constraints to the values of this coupling compatible with observations. However, we have assumed a chaotic inflationary model with potential $m_{\phi}^2 \phi^2$. It is expected that inflationary models with lower inflaton amplitudes during preheating will widen this range of values, as the value of the Ricci scalar $|R(t)|$ decreases, and hence the excitation of the Higgs field due to the tachyonic resonance is less strong. If the Standard Model potential does not have a second negative-energy vacuum at high energies, we cannot find upper bounds for $\xi_c$ in this way.

In this work, we have neglected the terms coming purely from the non-Abelian structure of the SM Lagrangian, considering instead that the $\lbrace W_a, Y \rbrace$ bosons can be regarded as Abelian gauge fields. We have argued that considering linear Abelian interactions leads to a larger excitation of the gauge fields, so that the non-Abelian terms can be safely ignored. We reach the important conclusion that the inclusion of gauge bosons in the system (even in the Abelian approach) does not change significantly the upper bound for $\xi$. The critical values $\xi_{c}^{(1),(2)}$ only change marginally when comparing both the absence and presence of gauge fields.

\acknowledgments
F.T. thanks the Theoretical Physics Group at Imperial College for kind hospitality during the development of this work. We acknowledge extensive use of the Imperial College High Performance Computing Service. F.T. is supported by the FPI-Severo Ochoa Ph.D. fellowship No. SVP-2013-067697. 
A.R. is supported by STFC grant ST/L00044x/1.
This work is also supported by the Research Project of the Spanish MINECO, Grant No. FPA2013-47986-03-3P, and the Centro de Excelencia Severo Ochoa Program No. SEV-2016-0597.

\appendix
 
\section*{Appendix: Lattice formulation}
\label{appendix}

\begin{figure}
      \begin{center} 
      \includegraphics[width=8.2cm]{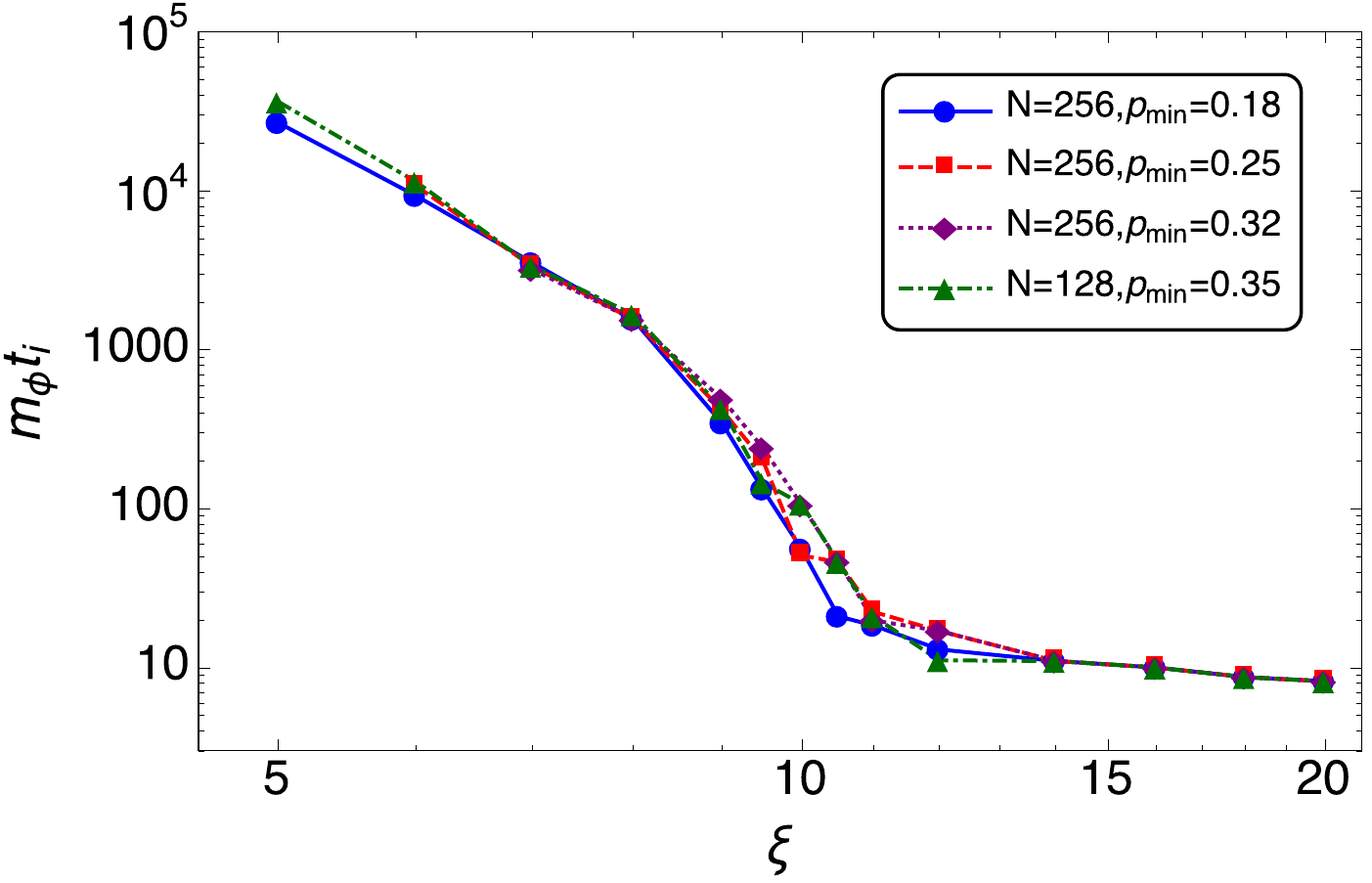}\hspace{0.5cm}
      \end{center} 
      \caption{We show the instability time $m_{\phi} t_i$ as a function of $\xi$ for $m_t=173.34 {\rm GeV}$, and  different lattice parameters. Each point corresponds to a single lattice simulation.}
      \label{fig:lattice-dependence}
\end{figure}

We present the discrete equations of motion we solve in our lattice simulations with gauge bosons, which as explained, include the four components of the Higgs field, and the three components of the effective gauge boson $A_{\mu}$, Eq.~(\ref{eq:super-boson}) (remember we set $A_0 = 0$). Let us define new spacetime variables as $(z, \vec{z} ) \equiv m_{\phi} (t, \vec{x})$, and rewrite the field variables in terms of dimensionless variables, $A_\mu \rightarrow A_\mu/m_\phi$, $\Phi \rightarrow \Phi/m_{\phi}$. Continuous action (\ref{eq:action-SM}) can be written in terms of these variables as
\bea
S &=& \int d^4 z \left\lbrace \frac{1}{4 g_s^2} (a \sum_i F_{0i}^2 - {1\over a} \sum_{i,j\neq i} F_{ij}^2 ) + a^3|D_0 \Phi |^2  \right. \nonumber \label{eq:app-contaction} \\
&& \left. - a \sum_i |D_i \Phi |^2 -  a^3 \left( 2\xi \mathfrak{R} |\Phi|^2 + \lambda(|\Phi|) |\Phi|^4 \right) \right\rbrace  \ , 
\eea
where we have defined $\mathfrak{R}$ as
\be \mathfrak{R} (t) \equiv 6 \left[ \left( \frac{a'}{a} \right)^2 + \frac{a''}{a} \right] = \frac{R(t)}{m_{\phi}^2} \ .
\ee
Let us consider a lattice cube with $N^3$ points, with time step ${\rm \dt} $ and lattice spacing ${\rm \dx}$. For a given field $f$, we define the following discrete forward and backward derivatives at a position $\vec{n}$ in the lattice as [$dx_0 \equiv {\rm dt}$, $dx_i \equiv {\rm dx }$, $(i=1,2,3)$]
\bea 
(\Delta_{\mu}^+ f) (\vec{n} ) &\equiv & \frac{1}{{\rm dx_{\mu}}} [ f (\vec{n} + \dx_{\mu} \hat{\sigma}_{\mu}) - f (\vec{n})] \ , \nonumber \\
(\Delta_{\mu}^- f) (\vec{n} ) &\equiv & \frac{1}{\rm dx_{\mu}} [ f (\vec{n}) - f (\vec{n} - \dx_{\mu} \hat{\sigma}_{\mu})] \ , \eea 
with $\hat{\sigma}_{\mu}$ a set of orthonormal vectors. Let us also define discrete lattice covariant derivatives as
\bea (D_{\mu}^+ f ) (\vec{n})& \equiv & \frac{1}{{\rm dx_{\mu}}} [ V_{\mu} f (\vec{n} + \dx_{\mu} \hat{\sigma}_{\mu} ) - f ( \vec{n} ) ] \ , \nonumber \\
( D_{\mu}^- f ) (\vec{n}) & \equiv & \frac{1}{{\rm dx_{\mu}}} [ f (\vec{n} ) - V_{-\mu}^* f ( \vec{n} - \dx_{\mu} \hat{\sigma}_{\mu}) ] \ ,\eea
where $V_{\mu} \equiv e^{-i A_{\mu} {\rm dx_{\mu}} }$ represents a `link'. We can write an action in the discrete equivalent to Eq.~(\ref{eq:app-contaction}) as
\be S= - \sum_{\hat{n}} {\rm \dt } {\rm \dx}^3  ( \mathcal{L}_{\rm I} + \mathcal{L}_{\rm II}  +  \mathcal{L}_{\rm III} + \mathcal{L}_{\rm IV}  + \mathcal{L}_{\rm V} ) \ , \label{eq:discreteaction}\ee
where the different pieces of the Lagrangian are
\bea
\mathcal{L}_{\rm I} &=& - \frac{a_{+0/2}}{2 g_s^2} \sum_i (\Delta_0^+ A_i - \Delta_i^+ A_0 )^2  \ , \nonumber \\
\mathcal{L}_{\rm II} &=& \frac{1}{ 4 g_s^2 a}  \sum_{i,j} (\Delta_i^+ A_j - \Delta_j^+ A_i)^2 \ ,\nonumber \\
\mathcal{L}_{\rm III} &=& - a_{+0/2}^3 (V_0 \Phi_{+0} - \Phi )^{\dagger} (V_0 \Phi_{+0} - \Phi ) \ ,  \nonumber \\
\mathcal{L}_{\rm IV} &=& a \sum_i (V_i \Phi_{+i} - \Phi )^{\dagger} (V_i \Phi_{+i} - \Phi ) \ , \nonumber \\
\mathcal{L}_{\rm V} &=& a^3 \left(2 \xi \mathfrak{R} |\Phi|^2 + \lambda (\varphi)|\Phi|^4\right) \ .
\eea
Here, we take the scale factor as evaluated at semi-integer times, and we define
\be a \equiv \frac{a_{+0/2} + a_{-0/2}}{2} \ . \ee

We can minimize this Lagrangian with respect $\varphi^{\dagger}$ and $A_i$ respectively, to obtain the discrete equations of motion. Setting $A_0 = 0$, they are
\bea 
&\Delta_0^- [a_{+0/2}^3 \Delta_0^+ \Phi ] - a \sum_i D_i^- D_i^+ \Phi  = \nonumber \\* 
& \hspace{2cm} - \left[ \frac{a^3}{2 | \Phi |} \frac{\partial V(|\Phi |)}{\partial |\Phi |} + \xi \mathfrak{R} \right] \Phi \ ,  \nonumber \\*
&\Delta_0^- (a_{+0/2} \Delta_0^+ A_i ) - \frac{1}{a} \sum_j\left[  \Delta_j^- \Delta_j^+ A_i - \Delta_j^- \Delta_i^+ A_j \right] = \nonumber \\*
&\hspace{3cm} - 2 g_s^2  \frac{a}{\dx} \mathfrak{Im} [ \Phi_{+i}^{\dagger} V_i \Phi ]
\label{eq:eom-gaugediscrete} \eea
On the other hand, minimizing the action with respect $A_0$, we obtain the Gauss conservation law in the discrete,
\be \sum_i \Delta_i^- \Delta_0^+ A_i = J_{\hat{n}} \ , \hspace{0.5cm} J_{\hat{n}} \equiv 2 g_s^2 \frac{a_{+0/2}^2}{ {\rm \dt }} \mathfrak{Im} [ \Phi_{+0}^{\dagger} \Phi ] \label{eq:gauss-discrete} \ . \ee
In order to trust the results from our lattice simulations, the fields must preserve then the condition
\be \Delta_G \equiv \frac{1}{N^3} \sum_{\hat{n}} \frac{\sum_i ( \Delta_i^- \Delta_0^+ A_i ) - J_{\hat{n}}   }{\sum_i ( \Delta_i^- \Delta_0^+ A_i ) + J_{\hat{n}} } \ll 1 \ .\ee
We have checked in our simulations that this is indeed the case, for all running time. Typically, $\Delta_G \lesssim 10^{-12}$.

The mininum and maximum momenta captured by the lattice are $p_{\rm min} = \frac{2 \pi}{L}$ and $p_{\rm max} = \frac{\sqrt{3} N}{2} p_{\rm min}$, with $L \equiv {\rm dx} N$. We must choose $N$ and $L$ so that we capture the relevant range of momenta of this system. Fig.~\ref{fig:lattice-dependence} shows the instability curve for the top quark mass $m_t=173.34 {\rm GeV}$, for different lattice parameters $p_{\rm min}$ and $N$. We see that the results are consistent, independently on the particular features of the lattice.

\bibliography{ReferencesV2}
\bibliographystyle{h-physrev4}

\end{document}